\documentclass[fleqn,usenatbib]{mnras}

\usepackage{mathptmx}

\usepackage[T1]{fontenc}

\DeclareRobustCommand{\VAN}[3]{#2}
\let\VANthebibliography\thebibliography
\def\thebibliography{\DeclareRobustCommand{\VAN}[3]{##3}\VANthebibliography}

\usepackage{graphicx}	
\usepackage{amsmath}	
\usepackage{amssymb}	
\usepackage{booktabs}
\usepackage[section]{placeins}
\usepackage[titletoc,title]{appendix}

\title[Temperatures of RSGs at low Z]{The temperatures of red supergiants in low metallicity environments}

\author[G. Gonz\'alez-Tor\`a et al.]{
Gemma Gonz\'alez-Tor\`a,$^{1}$\thanks{E-mail: G.GonzalezITora@2019.ljmu.ac.uk}
Ben Davies,$^{1}$
Rolf-Peter Kudritzki$^{2,3}$
and Bertrand Plez$^{4}$
\\
$^{1}$Astrophysics Research Institute, Liverpool John Moores University, Liverpool Science Park IC2, 
146 Brownlow Hill, Liverpool, L3 5RF, UK\\
$^{2}$LMU M\"unchen, Universit\"atssternwarte, Scheinerst. 1, 81679 M\"unchen, Germany\\
$^{3}$Institute for Astronomy, University of Hawaii at Manoa, 2580 Woodlawn Drive, Honolulu, HI 96822, USA\\
$^{4}$LUPM, UMR 5299, Universit\'e de Montpellier, CNRS, 34095 Montpellier, France
}

\date{Accepted 1 June 2021. Received 1 June 2021; in original form 9 February 2021}

\pubyear{2021}

\begin{document}
\label{firstpage}
\pagerange{\pageref{firstpage}--\pageref{lastpage}}
\maketitle


\begin{abstract}
The temperatures of red supergiants (RSGs) are expected to depend on metallicity (Z) in such a way that lower-Z RSGs are warmer.  
In this work, we investigate the Z-dependence of the Hayashi limit by analysing RSGs in the low-Z galaxy Wolf-Lundmark-Mellote (WLM), and compare with the RSGs in the  higher-Z environments of the Small Magellanic Cloud (SMC) and Large Magellanic Cloud (LMC).
We determine the effective temperature ($T_{\textrm{eff}}$) of each star by fitting their spectral energy distributions, as observed by VLT+SHOOTER, with MARCS model atmospheres. 
We find average temperatures of $T_{\textrm{eff}_{\textrm{WLM}}}=4400\pm202$ K, $T_{\textrm{eff}_{\textrm{SMC}}}=4130\pm103$ K, and $T_{\textrm{eff}_{\textrm{LMC}}}=4140\pm148$ K. From population synthesis analysis, we find that although the Geneva evolutionary models reproduce this trend qualitatively, the RSGs in these models are systematically too cool. 
We speculate that our results can be explained by the inapplicability of the standard solar mixing length to RSGs. 

\end{abstract}




\section{Introduction}

\label{ch:intro}
Red supergiants (RSGs) are an evolved state of massive stars. They have luminosities of $\sim 10^{4.39-5.52}\,L_{\odot}$ \citep{2020MNRAS.493..468D}, and comparing with evolutionary models  (e.g. \citealt{2013A&A...558A.103G}), a mass range of $\sim 8-25\,M_{\odot}$. The effective temperature ($T_{\textrm{eff}}$) of RSGs is thought to be a manifestation of the Hayashi limit. The Hayashi limit \citep{1961PASJ...13..442H} fixes the minimum $T_{\textrm{eff}}$ of the star and its maximum radius, where the star is known to have its most extended convective region and still maintaining hydrostatic equilibrium. 
Theoretical predictions show a dependency of the average $T_{\textrm{eff}}$ of RSGs on their metallicity (Z) (see for instance the series \citet{2000A&A...361..101M,2001A&A...373..555M,2002A&A...390..561M}, where stellar evolution in low metallicities has been extensively studied). 

An explanation for the $Z-T_{\textrm{eff}}$ dependency of RSGs from a physical point of view was proposed by \citet{2001A&A...373..555M}: for lower Z, the star has a lower opacity and molecular weight. This will result in a more compact main-sequence star with a smaller, hotter convective core. At the end of the main sequence, the He-burning core has a shallower potential, which corresponds to a smaller and warmer RSG. The prediction can be tested by studying the average $T_{\textrm{eff}}$ of RSGs in galaxies with different metallicities.

However, accurately measuring the $T_{\textrm{eff}}$ of RSGs is known to be problematic.   Specifically, \citet{2013ApJ...767....3D} showed that the $T_{\textrm{eff}}$ obtained from model fits to the TiO bands (which is the most prominent spectral feature for these cool stars) were systematically cooler than fits of the same models to the line-free continuum of the spectral energy distribution (SED). An explanation for the different $T_{\textrm{eff}}$ obtained when studying the TiO bands and the line-free continuum is provided by \citet{2013ApJ...767....3D}. Broadly, the line-free regions in the spectrum form close to the photosphere, where the local temperature is close to the $T_{\textrm{eff}}$ (Davies \& Plez 2021, accepted with minor revision). In contrast, the TiO bands can form much further out above the photosphere, where the temperature is $\sim 1000$ K lower.  To get consistent temperatures from both diagnostics, the model atmosphere used would have to be able to correctly predict the temperature structure from the photosphere out to many stellar radii. \citet{2013ApJ...767....3D} argued that 1D models provide a poor description of the outer temperature structure of RSGs. They suggested that models properly accounting for 3D radiative-hydrodynamics effects could yield consistent temperatures from both diagnostics. By adding a wind to the hydrostatic atmosphere, which increases the TiO absorption whilst leaving the rest of the optical-IR spectrum largely unchanged, a simultaneous fit to the TiO bands and the line-free continuum may be obtained (Davies \& Plez 2021, accepted with minor revision).

In this work, we use a $T_{\textrm{eff}}$  determination method independent of the TiO bands, by fitting the whole of the SED except the regions dominated by the molecular and line absorption (as in \citealt{2013ApJ...767....3D}). We have analysed three samples of RSGs  from the Local Group neighbouring galaxies Wolf-Lundmark-Mellote (WLM), Small Magellanic Cloud (SMC) and Large Magellanic Cloud (LMC), to investigate the Z-dependence of the average RSG $T_{\textrm{eff}}$ across a broader baseline than was previously studied in \citet{2013ApJ...767....3D}. The $T_{\textrm{eff}}$ and extinctions ($A_{\textrm{V}}$) have been obtained, as well as the bolometric luminosities ($L_{\textrm{bol}}$), this latter for WLM targets using synthetic and observed photometry. 
We also perform a population synthesis using the theoretical evolutionary tracks by \citet{2012A&A...537A.146E,2013A&A...558A.103G, 2019A&A...627A..24G}, and compare the observations with the predictions from the simulations to better study the $T_{\textrm{eff}}-\textrm{[Z]}$ dependency (where $[Z]=\log(Z/Z_{\odot})$, with respect to solar $Z_{\odot}=0.014$). 

\medskip

This paper is organised as follows: chapter~\ref{ch:datared} describes the observations and data reduction. The method used is presented in chapter~\ref{ch:analysis}, followed by the results obtained in chapter~\ref{ch:results} and a discussion on chapter~\ref{ch:discussion}, where we explore the implications of the determined temperatures and compare to simulations. Lastly, we conclude in chapter~\ref{ch:conclusion}. 

\section{Observations and Data reduction}
\label{ch:datared}

\subsection{New data}
We have observed several RSGs in WLM, selecting our targets from \citet{2012AJ....144....2L}. We chose the 9 brightest objects in the near-IR, which represents all RSGs in this galaxy with luminosities $\log(L_{\textrm{bol}}/L_{\odot}) > 4.4$ (see Fig.~\ref{fig:hrdiagram}). Our targets are listed in Table~\ref{tab:resultswlm}. We observed each star with VLT+XSHOOTER \citep{2006SPIE.6269E..33D} in order to obtain contemporaneous spectrophotometry from the optical to the near-IR, under the ESO program number (093.D-0021(A), PI: B Davies). All stars were observed with the 5\arcsec\ slit to minimise slit losses, in an ABBA nodding pattern. Slit positions were defined specifically to avoid any nearby stars clashing in the dispersion direction. The total integration times were the same for each star; 2248sec, 2760sec, and 3040sec in the UVB, VIS, and NIR arms respectively. The NIR integrations were broken up into discrete integration times of 190sec to avoid saturation in the airglow emission lines. In addition to the science targets, telluric standard stars of spectral type B were observed within 1.5hrs of any science exposure. Data were reduced following the same procedure described in \citet{2013ApJ...767....3D}.

When observing the stars from a very distant galaxy such as WLM, we cannot rule out the possibility that the RSGs observed are part of multiple systems, and other stars can contribute to the flux measured. However, at the resolution of ground-based survey imaging, we see no evidence of source confusion from abnormal colors or point spread functions. In addition, we see no evidence of hybrid spectral features (e.g. Balmer lines) in the blue, which would be indicative of an unresolved multiple system. Even with other targets being in the slit, RSGs are much brighter than anything else so the likelihood of significant contamination is small. 

While performing the analysis, we found that the WLM star number 7 (WLM 07) in Table~\ref{tab:resultswlm} has a radial velocity $v=30\,\textrm{km}/\textrm{s}$, while all the others are $v\sim120\,\textrm{km}/\textrm{s}$ . Checking their parallaxes at SIMBAD \citep{2000A&AS..143....9W}, WLM 07 has $1.4307 \pm 0.6966$ mas \citep{2018A&A...616A...1G}, while the other targets have parallaxes consistent with $0$ mas \citep{2018A&A...616A...1G}. Both velocities and parallaxes of WLM 07 are not coincident with the rest of the WLM targets in within the errors. Furthermore, this star has a spectral type (M3) which is much later than the others (K0-3, see Table~\ref{tab:resultswlm}). We have assumed that the target is a foreground star and not part of WLM. For these reasons, WLM 07 has been excluded from further analysis.

\subsection{Archival data}
For the LMC and SMC, we used the previous data from the VLT+XSHOOTER observations, under the ESO programme number 088.B-0014(A) (PI B. Davies). The observations, selection criteria and reduction steps are described in \citet{2013ApJ...767....3D,2015ApJ...806...21D}. The stars from LMC and SMC were selected from \citet{2006ApJ...645.1102L} to sample the full distribution of spectral types in each galaxy (as explained in \citealt{2013ApJ...767....3D}). 
As a further test that our sub-sample of stars in each galaxy has a distribution of $T_{\textrm{eff}}$ representative of the entire RSG population of that galaxy, we perform the following tests: we randomly select 10 RSGs from \citet{2018MNRAS.476.3106T} that have the same spectral type distribution as our sample stars. For each of these 10 stars, we randomly assign a $T_{\textrm{eff}}$ based on that star $T_{\textrm{eff}}$ measurement and associated error in \citet{2018MNRAS.476.3106T}. Next, we obtain the average $T_{\textrm{eff}}$  of these 10 randomly selected temperatures. We repeat this process 100 times for both LMC and SMC targets. Finally, we compare the results with the average $T_{\textrm{eff}}$ of the whole sample in \citet{2018MNRAS.476.3106T}. The error in the average temperatures is calculated with the mean error of the individual stars and the standard deviation.
\begin{table}
\centering
\begin{tabular}{c|c|c}

      Galaxy & Sub-sample $\overline{T}_{\textrm{eff}}$ (K)& Full sample $\overline{T}_{\textrm{eff}}$ (K) \\
     \hline
      LMC & $3800\pm50$ & $3810\pm100$ \\
      SMC & $3940\pm30$ & $3970\pm70$ \\
     \hline
     
\end{tabular}
\caption{Results of the statistical significance test. The left column indicates the galaxy studied, the middle column shows the average $T_{\textrm{eff}}$ retrieved given our sub-sample from \citet{2018MNRAS.476.3106T}, the right column shows the average $T_{\textrm{eff}}$ for the full sample in \citet{2018MNRAS.476.3106T} .\label{tab:stat}}
\end{table}

Checking the results in Table~\ref{tab:stat}, we see that the mean value for the sub-sample distribution is within the error limits of the average temperature of the whole RSGs sample in \citet{2018MNRAS.476.3106T}.
Both results are coincident within the error limits. Therefore, we conclude that our two sub-samples of stars have $T_{\textrm{eff}}$ distributions consistent with those of the entire population in each galaxy. 

\section{Determination of effective temperatures}
\label{ch:analysis}

We begin with a grid of model atmospheres generated with the MARCS code \citep{2008A&A...486..951G}. The 1D code assumes local thermodynamic equilibrium (LTE), hydrostatic equilibrium, and spherical symmetry. We fixed the metallicities to be $[Z]_{\textrm{WLM}}=-1.0$  \citep{2008ApJ...684..118U}, $[Z]_{\textrm{LMC}}=-0.35$, and $[Z]_{\textrm{SMC}}=-0.55$ \citep{2015ApJ...806...21D}. As for the metallicity adopted in WLM, in \citet{2008ApJ...684..118U} it is stated that previous photometric studies show WLM as having a young population in the disk, and an old metal-poor halo. By observing the red giant branch color, \citet{2005MNRAS.356..979M} found a metallicity of $[Z]=-1.45$ with respect to solar. This result however did not take into account the contribution of the younger population. When inspecting rich-metal line spectra of A supergiants, \citet{2008ApJ...684..118U} found an average metallicity of $[Z]=-0.87\pm0.06$. The $[Z]_{\textrm{WLM}}=-1.0$ adopted in this work is consistent with that found from analysis of WLM blue supergiants by \citet{2008ApJ...684..118U} once the differences in the Solar abundances used in that study and ours are taken into account (see Appendix of \citealt{2017ApJ...847..112D}). 

We assumed the fiducial values for the microturbulence of $\xi=3\,\textrm{kms}^{-1}$ (see  \citealt{2015ApJ...806...21D}). Surface gravities used are $\log\,\textrm{g}=-0.2$ for the Magellanic Clouds (MCs), as in \citet{2015ApJ...806...21D}, while for WLM $\log\,\textrm{g}=0$. The $\log\,\textrm{g}$ of the WLM stars was found iteratively, by comparing their luminosities and temperatures determined from our analysis with the evolutionary tracks of \citet{2013A&A...558A.103G}. We have found gravities ranging from $-0.2$ to $+0.4$, with the majority around $\sim 0.0$. The robustness of our results to these fiducial values for gravity, metallicity and microturbulence are discussed in Section~\ref{sec:robust}. The model grid is computed between $3400\,\textrm{K}<T_{\textrm{eff}}<5000\,\textrm{K}$ in steps of 100 K, which we then interpolate into a finer grid of 20 K steps. 

In addition to fitting for the best $T_{\textrm{eff}}$, we also allow the extinction $A_{\textrm{V}}$ to vary, since this parameter can make the star appear cooler and create a large degree of degeneracy with $T_{\textrm{eff}}$. For the MCs we use the extinction law of \citet{2003ApJ...594..279G}, which is specifically tuned to the interstellar medium in the direction of these galaxies. For WLM we use instead the law derived by \citet{1989ApJ...345..245C}, and assume a $R_{\textrm{V}}=3.1$ as of the Milky Way. The robustness of this assumption is explored in Section~\ref{sec:robust}. For the baseline (i.e. Galactic) extinction towards WLM, \citet{2014AAS...22311604S} shows that it is very low, and consistent with zero ($0\leq E(B-V) \leq 0.1$). Therefore, it will not influence our results. The grid of extinction parameters used is $0<A_{\textrm{V}}<2$ with a step of $0.01$.

\medskip
For the analysis, we fit the SED windows unaffected by line and molecular absorption as seen in Table~\ref{tab:cont}, which are the same used in \citet{2013ApJ...767....3D}. We avoid the \textit{BVR} spectral region, the molecular absorption bands of TiO, VO, CO (at $\sim\,1.5\,\mu$m and $>2.3\,\mu$m), and the CN band at $1.1\,\mu$ m. While \citet{2013ApJ...767....3D} performed a pixel-by-pixel matching of these spectral regions, we fit the mean flux of each SED window. The reason is that the SNR in the near-IR region of the WLM spectra are not high enough to do pixel-by-pixel analysis.

\begin{table}
\centering
\begin{tabular}{c|c}

      $\lambda_{min}$ (nm)& $\lambda_{max}$ (nm) \\
     \hline
      821.6 & 831.9 \\
      871.7 & 880.0 \\
      1036 & 1080 \\
      1212 & 1278 \\
      1610 & 1614 \\
      1649 & 1659 \\
      1716 & 1723 \\
      2120 & 2250 \\
     \hline
     
\end{tabular}
\caption{Regions used of the SED for the analysis.\label{tab:cont}}
\end{table}

At each point in the model grid (i.e. for each value of $T_{\textrm{eff}}$ and $A_{\textrm{V}}$) we adjusted the SED to the same data flux by minimising the $\sum(\log(SED_{\textrm{model}})-\log(SED_{\textrm{data}}))$ at the line-free continuum, where $\log(SED)$ represents the base-10 logarithm of the SED flux regions, in order to compare both fluxes.. We then performed a 2-parameter fit by means of a $\chi^{2}$ minimisation, assuming Gaussian correlation as shown in eq.~\ref{eq:chi2}: 

\begin{equation}
    \label{eq:chi2}
    \chi_{i}^{2}=\sum^{n}_{i=1}\frac{(SED_{\textrm{data}}-SED_{\textrm{model}})^{2}_{i}}{\sigma^{2}_{i}},
\end{equation}
\noindent where $n$ is the number of spectral regions in Table~\ref{tab:cont}, $SED_{\textrm{data}}$ represents the mean flux of the data for each one of the regions,  $SED_{\textrm{model}}$ the mean flux of the model, and $\sigma^{2}_{i}$ the standard deviation of the data in each SED region. 

The result of the analysis is a 2D-array of $\chi^{2}$, one value for each $A_{\textrm{V}}$ and $T_{\textrm{eff}}$. The best fit parameters correspond to those at the minimum value of the $\chi^{2}$ 2D-array. 

For the errors on $T_{\textrm{eff}}$ and $A_{\textrm{V}}$, we first determine the 68\% dispersion contours of the $\chi^{2}$ fit, that for two degrees of freedom like the present case ($T_{\textrm{eff}}$ and $A_{\textrm{V}}$), corresponds to all the grid models with values $\chi^{2}<\chi^{2}_{min}+2.3$ \citep{1976ApJ...210..642A}, where $\chi^{2}_{min}$ is the best-fitted model. The errors on each parameter are then defined as being the minimum values within this range.

\medskip
The $L_{\textrm{bol}}$ were calculated for WLM targets using synthetic photometry and available photometry at the SIMBAD database \citep{2000A&AS..143....9W}, with 2MASS \citep{2010PASP..122.1437P}, Gaia DR2 \citep{2018A&A...616A...1G}, PAN-STARSS \citep{2016arXiv161205560C}, VISTA \citep{2013Msngr.154...35M}, SkyMapper \citep{2018PASA...35...10W} and Spitzer/IRAC \citep{2015ApJS..216...10B} photometry. The magnitudes for the different filters were converted to flux, dereddened using the previously determined extinction $A_{\textrm{V}}$, and integrated over the wavelength range for all the photometric filters available (from $0.36\,\mu$m to $7.5\,\mu$m). We transformed to bolometric luminosities using the most recent distance determination $d_{\textrm{WLM}}=995\pm46\,\textrm{kpc}$ \citep{2008ApJ...684..118U}. \citet{2016A&A...592A..16D} showed that SMC RSGs display a high degree of spectral variability, finding evidence that variability increases with decreasing metallicity. Furthermore, \citet{2021arXiv210302609B} studied the effect of the variability between the minimum and maximum states of the RSGs in the stellar cluster Westerlund 1. Even when the most extreme assumptions were made, the resulting impact on the $L_{\textrm{bol}}$ was at most $\pm0.2$ dex. Therefore, any systematic uncertainty due to spectral variability in the stars in this work must be less than this.

To determine the error on $L_{\textrm{bol}}$, we propagated the errors through those on the individual flux measurements, the error on $A_{\textrm{V}}$, as well as the distance of the galaxy. Of these, the dominant source of uncertainty is the $A_{\textrm{V}}$.

\section{Results}
\label{ch:results}
\begin{figure*}
\centering
\includegraphics[width=.9\textwidth]{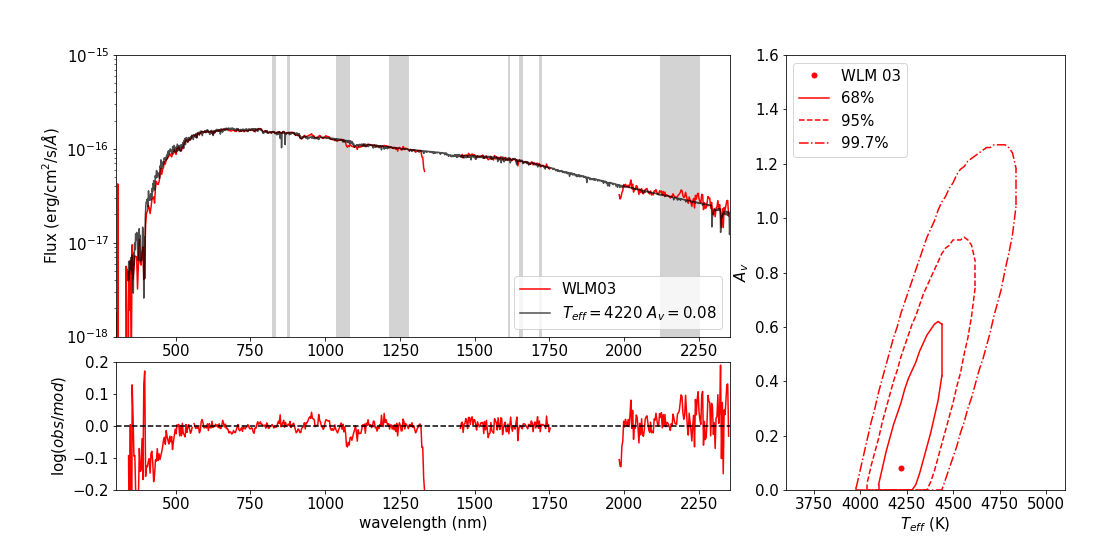}
\caption{Three panels showing the results of the analysis. \textit{Upper left:} Shows the smoothed data (red) and best fitted MARCS model (black), while the SED regions for the analysis are shown in gray. \textit{Lower left:} The residuals of the fit. \textit{Right:} shows the 68\%, 95\% and 99.7\% confidence limits for the best fitted $T_{\textrm{eff}}$ and $A_{\textrm{V}}$. }
\label{fig:wlmcase}
\end{figure*}

The best $T_{\textrm{eff}}$, $A_{\textrm{V}}$ and $\log(L_{\textrm{bol}}/L_{\odot})$ obtained after the analysis, for each object in WLM is listed in Table~\ref{tab:resultswlm} (with the errors from the 68\% $\Delta \chi^2$ isocontours), as well as their stellar coordinates, apparent magnitudes, spectral types from \citet{2012AJ....144....2L}, and identification names as in the SIMBAD database \citep{2000A&AS..143....9W}. Figure~\ref{fig:wlmcase} shows the results of the analysis for one case, where we see the best fit MARCS model (upper left panel) in black with respect to the data in red, the residuals of the fit in the lower left panel. The right panel of Figure~\ref{fig:wlmcase} shows the best fitted parameters along with the ellipse dispersion contours for the 68\% confidence limit, 95\% and 99.7\%. The best fits for the rest of the targets can be seen in Appendix~\ref{app:ap1}. 

In Appendix~\ref{app:ap1}, we see that for WLM 01, 05 and 09 there is a continuum shift for the long-wavelength edge in the observations. This is because in the last order of the NIR arm there was a variable thermal background which affected the flux calibration. To investigate the impact of this effect, we repeated the analysis shifting the fitted region at this order for these particular cases. The results did not change for the first 2 cases, and changed by $-20$ K in WLM 09. This does not increase the errors for these stars. Moreover, WLM 01 has an extinction close to zero of $A_{\textrm{v}}=0.00\pm1.02$, but one of the biggest error budgets of all WLM stars.
 
\begin{table*}
\centering

\begin{tabular}{c|c|c|c|c|c|c|c|c}

     Star & ID & RA & DEC & $m_{\textrm{R}}$ & $T_{\textrm{eff}}$ (K) & $A_{\textrm{V}}$ & $\log(L_{\textrm{bol}}/L_{\odot})$ & SpT \\
	\hline
	 \rule{0pt}{0.5cm}
	WLM 01 & LGGS J000153.17-152813.4 & 00 01 53.181 & -15 28 13.92 & 18.51 & $4580^{+420}_{-180}$ & $0.00^{+1.02}$ & $4.47^{+0.08}_{-0.12}$ & K0-1I \\
 \rule{0pt}{0.5cm}
	WLM 02 & LGGS J000156.77-152839.6 & 00 01 56.785 & -15 28 40.18 & 16.62 & $4660^{+340}_{-440}$ & $0.95^{+0.54}_{-0.95}$ & $5.52^{+0.07}_{-0.06}$ & K2-3I \\
 \rule{0pt}{0.5cm}
	WLM 03 & LGGS J000156.87-153122.3 & 00 01 56.887 & -15 31 22.84 & 17.91 & $4220^{+220}_{-120}$ & $0.08^{+0.54}_{-0.08}$ & $4.80^{+0.11}_{-0.05}$ & K0-1I \\
 \rule{0pt}{0.5cm}
	WLM 04 & LGGS J000157.01-152954.0 &  00 01 57.023 & -15 29 54.59 & 17.85 & $4560^{+440}_{-360}$ & $0.65^{+0.84}_{-0.65}$ & $4.89^{+0.09}_{-0.07}$ & K0-1I \\
 \rule{0pt}{0.5cm}
	WLM 05 & LGGS J000157.55-152915.8 & 00 01 57.545 & -15 29 16.05 & 18.41 &  $4160^{+780}_{-140}$ & $0.11^{+1.38}_{-0.11}$ & $4.66^{+0.13}_{-0.06}$ & K0-1I \\
 \rule{0pt}{0.5cm}
	WLM 07 & LGGS J000158.14-152332.2 & 00 01 58.146 & -15 23 32.58 & 18.62 &  $4460^{+540}_{-140}$ & $0.00^{+1.15}_{-0.01}$ & $4.48^{+0.09}_{-0.04}$ & M3I \\
 \rule{0pt}{0.5cm}
	WLM 08 & LGGS J000158.74-152245.5 & 00 01 58.746 & -15 22 46.03 & 17.56 &  $4420^{+560}_{-220}$ & $0.40^{+1.04}_{-0.40}$ & $4.98^{+0.07}_{-0.06}$ & K0-1I \\
 \rule{0pt}{0.5cm}
	WLM 09 & LGGS J000200.81-153115.7 & 00 01 59.610 & -15 30 59.90 & 18.07 & $4340^{+160}_{-160}$ & $0.34^{+0.41}_{-0.34}$ & $4.80^{+0.08}_{-0.07}$ & K2-3I \\
 \rule{0pt}{0.5cm}
	WLM 10 & LGGS J000200.81-153115.7 & 00 02 00.810 & -15 31 15.70 & 17.78 &  $4380^{+200}_{-220}$ & $0.60^{+0.51}_{-0.57}$ & $4.84^{+0.10}_{-0.06}$ & K0-1I \\

     \hline     
\end{tabular}
\caption{Table shows the ID name, RA, DEC, $m_{\textrm{R}}$ (in magnitudes), spectral type as indicated at the SIMBAD database, and the best $T_{\textrm{eff}}$, $A_{\textrm{V}}$ and $\log(L_{\textrm{bol}}/L_{\odot})$ obtained for each studied RSG at the WLM galaxy, within their 68\% confidence limits. WLM 07 is suspected to be a foreground star. \label{tab:resultswlm}}
\end{table*}

\begin{figure*}
\centering
\includegraphics[width=.9\linewidth]{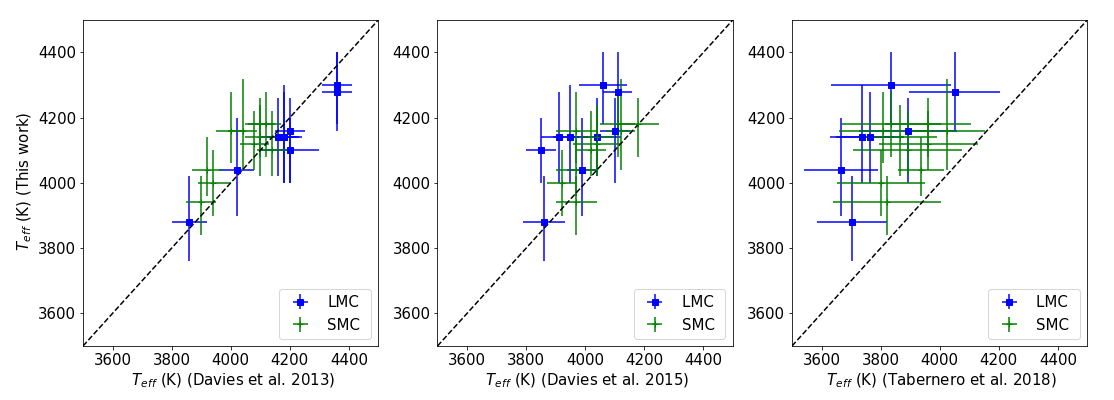}
\caption{Comparison between the results from this work (y-axis) and the previous results found by \citet{2013ApJ...767....3D} (\textit{left panel}), \citet{2015ApJ...806...21D} (\textit{center panel}) and \citet{2018MNRAS.476.3106T} (\textit{right panel}). The SMC targets are represented in green, while the LMC in blue squares. The black-dashed line shows the 1:1 ratio.}
\label{fig:mccomp}
\end{figure*}

Table~\ref{tab:resultsmc} shows the calculated values with the SED method for the SMC and LMC stars, with the errors in $T_{\textrm{eff}}$ and $A_{\textrm{V}}$ from the 68\% $\Delta \chi^2$ isocontours. We have compared the stars with the previous results from \citet{2013ApJ...767....3D,2015ApJ...806...21D,2018MNRAS.476.3106T} in Figure~\ref{fig:mccomp}. The  results are consistent between $1-2\sigma$ for the first two panels, the discrepancy with \citet{2018MNRAS.476.3106T} is discussed in Section~\ref{sec:comp}.


\begin{table}
\centering
\begin{tabular}{c|c|c}
  Star & $T_{\textrm{eff}}$ (K) & $A_{\textrm{V}}$ \\
  \hline
\hline
   \rule{0pt}{0.5cm}
SMC 011709 & $4100^{+120}_{-80}$ & $0.49^{+0.35}_{-0.3}$ \\
   \rule{0pt}{0.5cm}
SMC 013740 & $4040^{+100}_{-80}$ & $0.81^{+0.32}_{-0.32}$ \\
   \rule{0pt}{0.5cm}
SMC 020133 & $4160^{+120}_{-100}$ & $1.27^{+0.37}_{-0.39}$ \\
   \rule{0pt}{0.5cm}
SMC 021362 & $3940^{+80}_{-100}$ & $0.37^{+0.33}_{-0.35}$ \\
   \rule{0pt}{0.5cm}
SMC 030616 & $4140^{+100}_{-120}$ & $0.73^{+0.35}_{-0.39}$ \\
   \rule{0pt}{0.5cm}
SMC 034158 & $4180^{+80}_{-100}$ & $0.55^{+0.32}_{-0.37}$ \\
   \rule{0pt}{0.5cm}
SMC 035445 & $4120^{+100}_{-80}$ & $0.46^{+0.29}_{-0.28}$ \\
   \rule{0pt}{0.5cm}
SMC 049478 & $4180^{+100}_{-100}$ & $0.75^{+0.34}_{-0.37}$ \\
   \rule{0pt}{0.5cm}
SMC 050840 & $4000^{+100}_{-100}$ & $0.55^{+0.33}_{-0.36}$ \\
   \rule{0pt}{0.5cm}
SMC 057386 & $4160^{+160}_{-120}$ & $0.39^{+0.44}_{-0.38}$ \\
   \rule{0pt}{0.5cm}
LMC 064048 & $3880^{+140}_{-120}$ & $0.52^{+0.61}_{-0.52}$ \\
   \rule{0pt}{0.5cm}
LMC 067982 & $4140^{+140}_{-140}$ & $1.52^{+0.47}_{-0.60}$ \\
   \rule{0pt}{0.5cm}
LMC 116895 & $4140^{+160}_{-140}$ & $1.10^{+0.65}_{-0.56}$ \\
   \rule{0pt}{0.5cm}
LMC 131735 & $4280^{+120}_{-120}$ & $0.81^{+0.50}_{-0.44}$ \\
   \rule{0pt}{0.5cm}
LMC 136042 & $4100^{+100}_{-100}$ & $1.99^{+0.01}_{-0.30}$ \\
   \rule{0pt}{0.5cm}
LMC 137818 & $4040^{+160}_{-140}$ & $0.85^{+0.53}_{-0.55}$ \\
   \rule{0pt}{0.5cm}
LMC 142202 & $4160^{+100}_{-160}$ & $1.79^{+0.20}_{-0.66}$ \\
   \rule{0pt}{0.5cm}
LMC 143877 & $4300^{+100}_{-120}$ & $1.22^{+0.47}_{-0.45}$ \\
   \rule{0pt}{0.5cm}
LMC 158317 & $4140^{+120}_{-120}$ & $1.55^{+0.44}_{-0.54}$ \\
\hline 
\end{tabular}
 \caption{Best $T_{\textrm{eff}}$ and $A_{\textrm{V}}$ obtained for each studied RSG at the SMC and LMC galaxies, within their 68\% confidence limits. The stars names are based on the catalogue available by \citet{2018yCat..74783138D}.\label{tab:resultsmc}} 
\end{table}

There is one particular case in LMC that needs to be taken with care. Checking Figure~\ref{fig:lmc5}, we see that for LMC 136042 the observations (upper left panel, in blue) tend to upper the continuum in the spectral region of <500 nm, meaning that their flux contribution at the green-blue region of the spectra is higher than expected. This indicates that there is a blue star next to this target \citep{2015A&A...578A...3G}. We excluded this target for the rest of the analysis. 

\begin{figure}
\centering
\includegraphics[width=1.\linewidth]{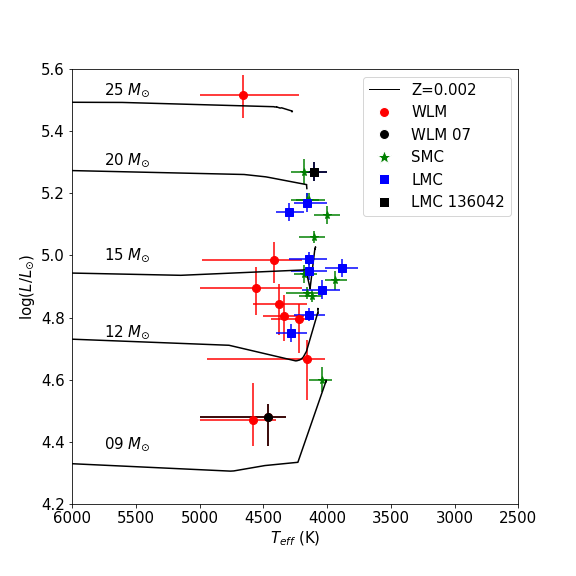}
\caption{HRD showing the RSGs of WLM (red), LMC (blue), SMC (green), with the evolutionary tracks from \citet{2013A&A...558A.103G}, with a $Z=0.002$, and rotation $v/v_{c}=0.4$ comparable to the galaxies studied. The black dots correspond to the problematic targets. The luminosities for the MCs are taken from \citet{2018yCat..74783138D}. }
\label{fig:hrdiagram}
\end{figure}

The stars are placed in the Hertzprung-Russel diagram (HRD) in Figure~\ref{fig:hrdiagram}, based on their $L_{\textrm{bol}}$ and $T_{\textrm{eff}}$. We have also plotted for comparison the $9-20\,M_{\odot}$ rotating evolutionary tracks for RSGs taken from \citet{2013A&A...558A.103G}, in  black solid-line showing the tracks corresponding to a $Z=0.002$, comparable to the studied galaxies. The average values for the three galaxies (excluding WLM 07 and LMC 136042, for the reasons already mentioned), are:
\begin{center}
 WLM: $T_{\textrm{eff}}=4400\pm202$ K, 
\medskip

 SMC: $T_{\textrm{eff}}=4130\pm103$ K, 
\medskip 

 LMC: $T_{\textrm{eff}}=4140\pm148$ K, 
\end{center}

\noindent where the errors correspond to the quadratic sum of the formal errors from the model fitting and the systematic errors arising from our assumptions when fitting the data (see Section~\ref{sec:robust}).

With the results above we see that, as theory predicts (e. g. \citealt{2001A&A...373..555M}) there is a dependence with average $T_{\textrm{eff}}$ of RSGs on Z for WLM and the MCs. Indeed the lowest-Z galaxy shows an average temperature $\sim 250$ K higher than the MCs. In the case of the MCs, we are not able to resolve a temperature difference between these two galaxies. This result agrees with previous work from \citet{2015ApJ...806...21D}, which claimed that there is no measurable dependence of $T_{\textrm{eff}}$ on $Z$ between the metallicities of LMC and SMC. In this current work we extend the metallicity baseline down to $[Z]=-1$, after which a trend on $T_{\textrm{eff}}$ with $Z$ becomes clear (see Figure~\ref{fig:simulations2}).

\subsection{Robustness of the analysis}
\label{sec:robust}
As described in Section~\ref{ch:analysis}, we made several assumptions in our fitting methodology. In this section, we now investigate the sensitivity of our results to these assumptions.

\subsubsection{Microturbulence}
To investigate the effect of varying our choice of $\xi$ for the models employed, we repeated the analysis for each target with $\xi=2$ and 4 km/s, since \citet{2015ApJ...806...21D} found a range of $\xi$ between $2-4$ km/s for all MCs stars. The median $T_{\textrm{eff}}$ does not change for WLM when assuming these $\xi$ values, while for the MCs it changes in both by $-10$ K with $\xi=2$, and $+20$ K for $\xi=4$. We see that the effect of varying $\xi$ is a factor of 10 smaller than the fitting errors. Therefore, a systematically different $\xi$ in WLM cannot explain the $T_{\textrm{eff}}$ differences between this galaxy and the MCs.

In short, the choice of $\xi$ makes little to no difference to our results. The reason is that $\xi$ will affect the strong absorption lines of the spectra, but in this work we are studying the line-free continuum.

\subsubsection{Surface gravity}
We also perform the analysis with $\log\,\textrm{g}=-0.5$ and $\log\,\textrm{g}=+0.5$ based on the gravity range found in \citet{2015ApJ...806...21D}. We find that for $\log\,\textrm{g}=-0.5$ in WLM the average temperature varies by $-100$ K while for the MCs it varies by $-40$ K for both. With $\log\,\textrm{g}=+0.5$ the variation is $+80$ K for WLM, $+30$ K for SMC, and $+40$ K for LMC. As a consequence, we see that the choice of surface gravity seems to make an impact on the final results in WLM, while for the MCs the variation is again smaller. As commented in Section~\ref{ch:analysis}, we expect the range on surface gravities to be $-0.2<\log\,\textrm{g}< +0.4$, with the majority around 0. This is comparable with the results from \citet{2015ApJ...806...21D}. 

To sum up, there is a stronger degeneracy when studying the effect of $\log\,\textrm{g}$. This offset seems to be of $\pm80$ K, but cannot explain why the $T_{\textrm{eff}}$ in WLM is $\sim250$ K higher than the MCs. 
\subsubsection{Extinction law}
To check the robustness of the extinction law for WLM, we have performed the analysis using various values of $R_{\textrm{V}}$ from 2 to 6. The results changed by less than 30 K for $R_{\textrm{V}}=2$, 4 and 5, while for a extreme assumption such as $R_{\textrm{V}}=6$ the difference was of only $+50$ K. Since WLM is a metal poor environment, we also studied the effect of using a SMC-like extinction law, and found a difference on the average temperature of  $-20$ K, all between the error limits. The best fit for $A_{\textrm{V}}$ changes a maximum of $+0.09$ for the extreme assumption of $R_{\textrm{V}}=6$, and it is within the error limits for $A_{\textrm{V}}$.

In conclusion, the choice of $R_{\textrm{v}}$ and extinction law makes barely any difference when determining the $T_{\textrm{eff}}$.

\subsubsection{SED continuum regions}
The last assumption that needs to be checked is the choice of the SED continuum regions. We have recalculated our results with slightly different SED regions (e. g. including a continuum region at $\sim400$ nm, or varying the \textit{JHK} band regions in Table~\ref{tab:cont} by $\sim 50$ nm). We find that the choice of continuum windows can shift the median $T_{\textrm{eff}}$ in each galaxy by $\pm 60$ K. We interpret this error to represent the absolute level of accuracy on our results. In other words, a median $T_{\textrm{eff}}$ difference between two galaxies within $60$ K would be consistent with zero. 

Lastly, the choice of the SED continuum regions give a systematics of $\sim60$ K, this is smaller than the limitations obtained by the 68\% dispersion contours, and can be related with the accuracy of the method.

\subsection{Comparison with previous studies}
\label{sec:comp}

\citet{2013ApJ...767....3D} used two different methods to calculate the $T_{\textrm{eff}}$ of the same targets: the first one has been conventionally used to analyse M-K type stars, and is based on fitting the TiO spectral bands (between 500 and 800 nm). The second method is the same SED continuum fit as used for this work, but with a $\xi=2$ km/s\footnote{As the \citet{2013ApJ...767....3D} study pre-dated the detailed spectral fitting in \citet{2015ApJ...806...21D} in which the microturbulent velocities were measured.}. 
In the left panel of Figure~\ref{fig:mccomp} we compare our results to those of \citet{2013ApJ...767....3D}. We notice that the star-to-star differences with  \citet{2013ApJ...767....3D} are always in between $1\sigma$ for both MCs, and there is no systematic offset. The very slight discrepancies between the results from this work (y-axis) and \citet{2013ApJ...767....3D} (x-axis) can be justified by the change in $\log\,g$ and $\xi$, as seen in Sect~\ref{sec:robust}. 

In \citet{2015ApJ...806...21D} the method consists of comparing the strengths of different spectral lines on the J-band with non-LTE model grids. We have used the same $\log\,g$ and $\xi$ as in \citet{2015ApJ...806...21D}. In the center panel of Figure~\ref{fig:mccomp}, we can see that the results are consistent between $2\sigma$, and obtain a median offset of $\sim60$ K, which is within the level of precision obtained by the choice of SED regions in Section~\ref{sec:robust}. Despite the two completely different methodologies employed, the agreement is excellent. 

The work of \citet{2018MNRAS.476.3106T} differs the most from this work (right panel in Figure~\ref{fig:mccomp}). We have calculated a median offset of $\sim 150$ K between both results, ours being warmer. In \citet{2018MNRAS.476.3106T}, temperatures are estimated by fitting spectral lines in the I-band with predictions from MARCS model atmospheres and LTE line formation.

The differences between our results and those of \citet{2018MNRAS.476.3106T} are twofold. Firstly, these LTE models do not include the non-LTE correction employed in \citet{2015ApJ...806...21D}. This can account for $\sim50$ K (see \citealt{2013ApJ...764..115B}), assuming that the corrections in the I-band are similar to those in the J-band.

\begin{figure}
\centering
\includegraphics[width=1.\linewidth]{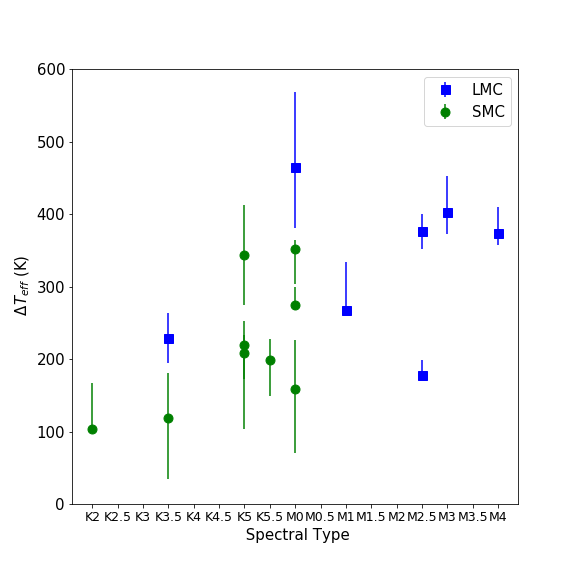}
\caption{Panel showing the difference between the $T_{\textrm{eff}}$ results from this work and \citet{2018MNRAS.476.3106T}, with respect to the spectral type of each RSG studied in the LMC (green dots) and the SMC (blue squares), as in the SIMBAD database.}
\label{fig:tab}
\end{figure}

Secondly, the region studied in \citet{2018MNRAS.476.3106T} of the I-band overlaps with a TiO absorption band. \citet{2013ApJ...767....3D} showed that the MARCS models cannot simultaneously reproduce the TiO bands and the continuum (see Figure set 1 in \citealt{2005ApJ...628..973L} and Figure set 2 in \citealt{2006ApJ...645.1102L}). This makes continuum placement in the I-band gradually more problematic at spectral types M0 and later, as the contamination by TiO grows. Therefore, we would expect to see a trend of increasing disparity between the $T_{\textrm{eff}}$ of \citet{2018MNRAS.476.3106T} and those of our study with increasing spectral type. Indeed, this is what we see in Figure~\ref{fig:tab}, a trend for the increasing difference between the $T_{\textrm{eff}}$ of \citet{2018MNRAS.476.3106T} and those of our work ($\Delta T_{\textrm{eff}}$) as we go to later spectral types, for both LMC and SMC. Therefore, we conclude that the systematic offset in Figure~\ref{fig:mccomp} can at least be partially explained by this trend.

\section{Discussion}
\label{ch:discussion}
In Section~\ref{ch:results}, we have found an average $T_{\textrm{eff}}$ of RSGs in WLM warmer than for the higher Z environments of the MCs. As seen in Section~\ref{sec:robust}, these differences cannot be explained by systematics in the fitting method. In this section, we investigate how our results compare to the expectations from stellar evolution models.

To better understand how our results compare to model predictions, we performed a population synthesis analysis. We generated a number of simulated stars ($\sim 10^{5}$) with masses of $5-100\,M_{\odot}$ according to a Salpeter initial mass function. Then we assigned random ages, and used the evolutionary tracks of \citet{2012A&A...537A.146E,2013A&A...558A.103G, 2019A&A...627A..24G} to interpolate their luminosities and $T_{\textrm{eff}}$ at that age. We disregarded  stars that are not in the RSG phase or are older than their expected lifetime. We have then selected the stars with $4.4<\log(L_{\textrm{bol}}/L_{\odot})<5.6$ (following \citealt{2020MNRAS.493..468D}) and $3000 < T_{\textrm{eff}} < 5000$ K, corresponding to a RSG phase and the stars in our sample. {We determine the $T_{\textrm{eff}}$ distribution of the stars in the RSG phase. Figure~\ref{fig:simulations} shows the histograms of the $T_{\textrm{eff}}$ from the RSGs after the population synthesis, for the case of non-rotating simulated stars with Z=0.014, 0.002 and 0.0004. As expected, we see that for higher metallicities there is a trend to lower temperatures. This trend does not change if we use a $\log(L_{\textrm{bol}}/L_{\odot})=4.5$ or $5$ cut instead. 

\begin{figure}
\centering
\includegraphics[width=1.\linewidth]{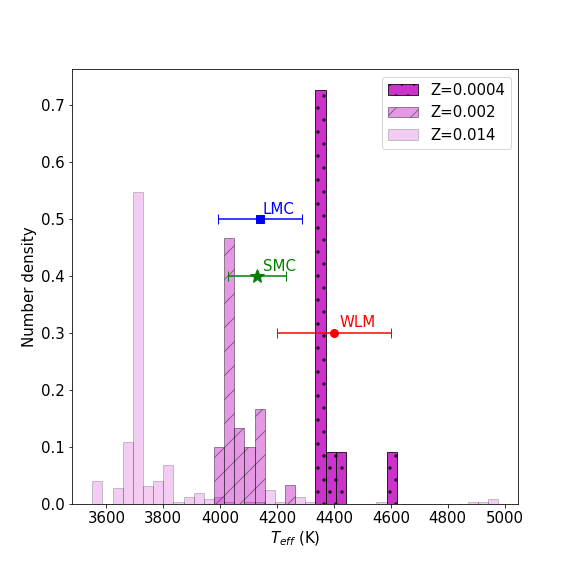}
\caption{Histograms showing the number density of simulated stars, in solid magenta bins for Z=0.014, dashed bins for Z=0.002 and spotted bins for Z=0.0004. The median temperature and error bars calculated with the observations for the three galaxies is also shown in green (SMC), blue (LMC) and red (WLM). The simulations shown are obtained for non-rotating stars.}
\label{fig:simulations}
\end{figure} 

A clearer visualisation of this Z dependence with $T_{\textrm{eff}}$ can be seen in Figure~\ref{fig:simulations2}: the median $T_{\textrm{eff}}$ of the simulated stars is shown with respect to their $[Z]$, in magenta for non-rotating and in yellow for rotating simulated stars. The shaded regions show the 68\% confidence limits (analogous to a $1\sigma$ standard deviation). The median $T_{\textrm{eff}}$ are also shown in Figure~\ref{fig:simulations2} for the observations of each galaxy. 
Comparing the model predictions with the data, we point out the following:
\begin{itemize}
\item For all three galaxies, the RSGs appear to be warmer than the model predictions.
\item The results show a similar qualitative slope in comparison with the simulations, but the models seem to be too cool, especially for the non-rotating models. This systematic offset is $\Delta T_{\textrm{eff}_\textrm{WLM}}=220\,\textrm{K}$, $\Delta T_{\textrm{eff}_\textrm{SMC}}=140\,\textrm{K}$, and $\Delta T_{\textrm{eff}_\textrm{LMC}}=250\,\textrm{K}$ for rotating models, while $\Delta T_{\textrm{eff}_\textrm{WLM}}=300\,\textrm{K}$, $\Delta T_{\textrm{eff}_\textrm{SMC}}=220\,\textrm{K}$, and $\Delta T_{\textrm{eff}_\textrm{LMC}}=300\,\textrm{K}$ for non-rotating. The offset is also out of the error dispersion for the LMC. The significance of this offset is discussed below.
\end{itemize}  

To determine the statistical significance of the offset of the observed $T_{\textrm{eff}}$ with respect to the model predictions, we perform a Monte-Carlo (MC) test. Each MC trial is constructed as follows: firstly, for each galaxy, we randomly select $N$ stars, where $N$ is the number of stars observed in that galaxy (e.g. $N=10$ in SMC). We note an individual star may be selected more than once per trial. Then, for each randomly selected star, we randomly assign a $T_{\textrm{eff}}$ from that star's observed probability distribution (illustrated by e. g. the right panel in Figure~\ref{fig:wlmcase}). Finally, we determine the median $T_{\textrm{eff}}$, $\overline{T}_{\textrm{eff}}$, of the $N$ stars. We then repeat this process 1000 times to determine the probability distribution on $\overline{T}_{\textrm{eff}}$ for each galaxy. 

Using Figure~\ref{fig:simulations2}, we interpolate the simulated $T_{\textrm{eff}}$ and its uncertainty at the metallicity of the galaxy, and its corresponding error limits.  
Next, we calculate the probability that the model fits the data by integrating the product of the observed and simulated $T_{\textrm{eff}}$ distributions. We repeat this process for every galaxy. 

For rotating models, we find a probability that the model and data agree of $p_{\textrm{WLM}}=0.07$, $p_{\textrm{SMC}}=0.03$, and $p_{\textrm{LMC}}=0.01$. The probability that all three galaxies are consistent with the rotating model predictions is $p_{\textrm{total}}=10^{-5}$. For non-rotating models we find $p_{\textrm{WLM}}=0.003$, $p_{\textrm{SMC}}=0.003$, $p_{\textrm{LMC}}=0.0004$, and $p_{\textrm{total}}=10^{-7}$. We conclude that the systematic offset between observed and predicted $T_{\textrm{eff}}$ in all galaxies in this study cannot be explained by random scattering within the experimental errors. 

\medskip
A possible explanation to this mismatch between observations and simulations could be a breakdown in the assumptions used to simulate convection in RSGs. In 1D models, the mixing length theory (MLT) \citep{1958ZA.....46..108B} is the analytic approximation used to describe the 3D phenomenon of convection. It assumes that a fluid parcel can travel a distance fixed by the so-called mixing length, $l$, before dispersing into the surrounding material. This mixing length is usually expressed in terms of the pressure scale height, $\alpha_{\textrm{MLT}}=l/H_{P}$, where $\alpha_{\textrm{MLT}}$ is the mixing length parameter. This free parameter is usually calibrated using the standard solar model with a single depth-independent $\alpha^{\odot}_{\textrm{MLT}}$. This solar calibrated value is used for stars of all masses, metallicities, and evolutionary phases. Indeed, the evolutionary tracks in this work by \citet{2012A&A...537A.146E,2013A&A...558A.103G, 2019A&A...627A..24G} use $\alpha_{\textrm{MLT}}=\alpha^{\odot}_{\textrm{MLT}}=1.6467$ for massive star models, arguing that $\alpha_{\textrm{MLT}}$ only changes to $1.6$ for very high mass stars, $M=150M_{\odot}$ \citep{2013A&A...558A.103G}. This variation was found by accounting for the differences in the massive stars equation of state. 

However, 3D simulations of convection in low mass stars have shown that there is a strong $\alpha_{\textrm{MLT}}$ dependency on $T_{\textrm{eff}}$ and $\log\,\textrm{g}$ (e.g.  \citet{2014MNRAS.445.4366T}, where for $T_{\textrm{eff}}<5000$ K a $\alpha_{\textrm{MLT}}>1.8$ is found). Work by \citet{2015A&A...573A..89M} also finds a difference of $\sim 20\%$ in the $\alpha_{\textrm{MLT}}$ depending on the mass of the star. Although these studies do not extend to the parameter ranges relevant for RSGs, they point out that the value of $\alpha_{\textrm{MLT}}$ is not independent of the mass.

Specifically studying RSGs, \citet{2018ApJ...853...79C} adopts the approach of tuning $\alpha_{\textrm{MLT}}$ to match the locations of RSGs in a range of metallicity environments. Regardless of which RSG temperature scales are used, they argue for a metallicity dependent $\alpha_{\textrm{MLT}}$. Moreover, \citet{2013MNRAS.433.1745D} studies the supernova type II-P (SN IIP) progenitors, varying the parameters (e.g. mixing length, overshoot, rotation, metallicity) in the stellar evolution code MESA STAR. They find that the RSG radii should be reduced in comparison with \citet{2005ApJ...628..973L}, implying that the $T_{\textrm{eff}}$ of RSGs should be higher than the \citet{2005ApJ...628..973L} temperature scale.

The previously described work, in combination with our results presented in this study, are part of a growing body of evidence that the assumption of a solar mixing length parameter is not adequate to explain the locations of RSGs in the HRD as a function of metallicity.

\begin{figure}
\centering
\includegraphics[width=1.\linewidth]{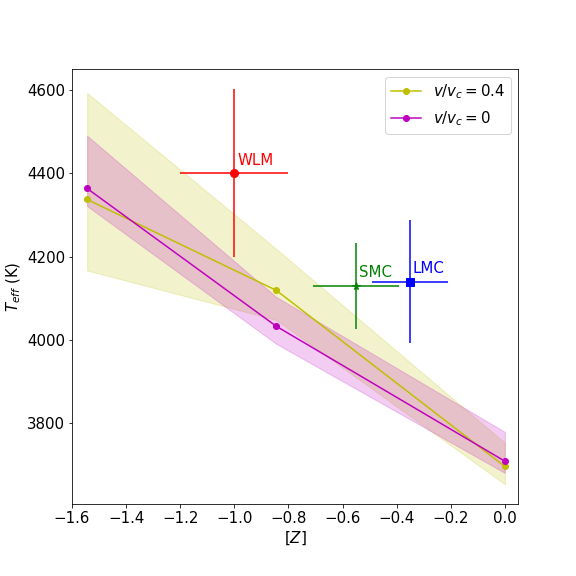}
\caption{Plot showing the temperature trend with respect to the metallicity ([Z]), in yellow for rotating simulated stars, and magenta for non-rotating. The error limits in the trend are calculated using the 68\% dispersion of the cumulative distribution for the skewed histograms. The median temperatures and error bars calculated with the observations for the three galaxies are also shown in green (SMC), blue (LMC) and red (WLM). }
\label{fig:simulations2}
\end{figure} 

\section{Conclusions}
\label{ch:conclusion}
We have analysed a total of 28 RSGs observed with VLT-XSHOOTER from the neighbouring galaxies LMC, SMC and WLM, by fitting the flux of the SED regions free from molecular features.
Our main conclusions are as follows:
\begin{itemize}
\item We find an average RSG $T_{\textrm{eff}}$ for WLM which is $\sim300$ K warmer than that in either of the MCs. This trend of increasing average $T_{\textrm{eff}}$ with decreasing Z is in qualitative agreement with theoretical predictions. 
 
\item From population synthesis analysis, we find that there is a systematic offset between expected and observed temperatures of RSGs at all metallicities. Specifically, RSGs in evolutionary models are too cool by $\sim200$ K. This could be due to a wrong estimation of the mixing length parameter for 1D models of massive stars.
 
\end{itemize}
 
\section*{Acknowledgements}
We would like to thank the referee for useful comments which helped improve the paper. GGT is supported by a scholarship from the Liverpool John Moores University. GGT wants to thank the useful comments from Ricardo Schiavon, Sarah McDonald, and Sylvia Ekstr\"om.
RPK acknowledges support by the Munich Excellence Cluster Origins funded by the Deutsche Forschungsgemeinschaft (DFG, German Research Foundation) under Germany's Excellence Strategy EXC-2094 390783311. 

\section*{Data availability}
The data underlying this article will be shared on reasonable request to the corresponding author. The datasets were derived from sources in the public domain: http://archive.eso.org/cms.html.


\bibliographystyle{mnras}
\bibliography{references} 





\begin{appendices}
\appendix
\section{Analysis of RSGs}
\label{app:ap1}
\begin{figure*}
\centering
\includegraphics[width=1.\linewidth]{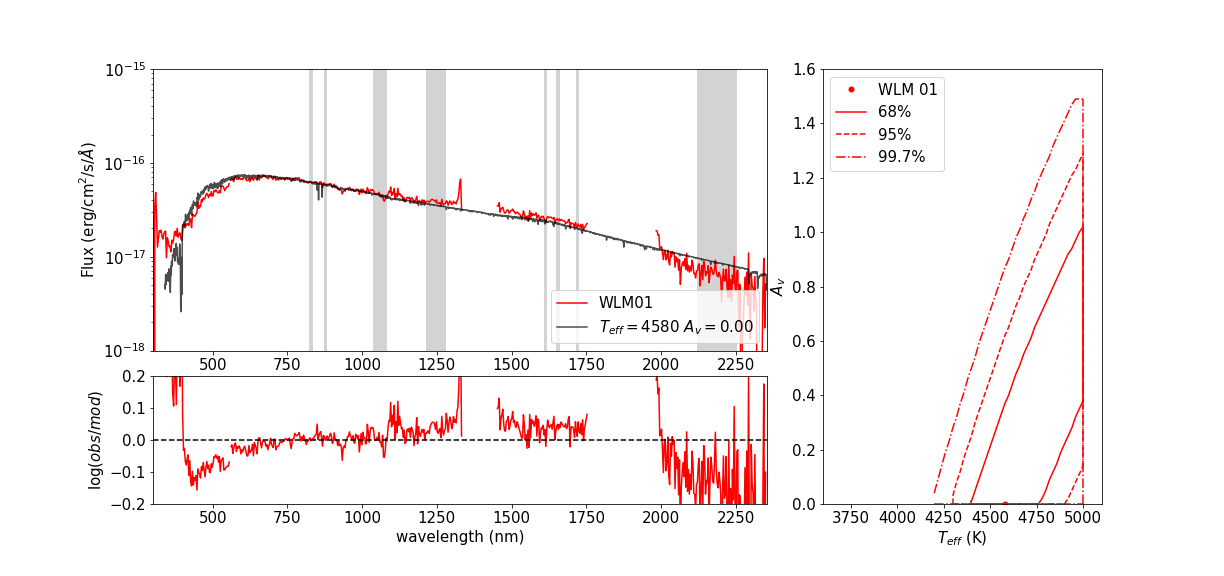}
\caption{Three panels showing the results of the analysis for WLM 01. \textit{Upper left:} Shows the smoothed data (red) and best fitted MARCS model (black), while the SED regions for the analysis are shown in gray. \textit{Lower left:} The residuals of the fit. \textit{Right:} shows the 68\%, 95\% and 99.7\% confidence contours for the best fitted effective temperature and extinction. }
\label{fig:wlm1}
\end{figure*}

\begin{figure*}
\centering
\includegraphics[width=1\linewidth]{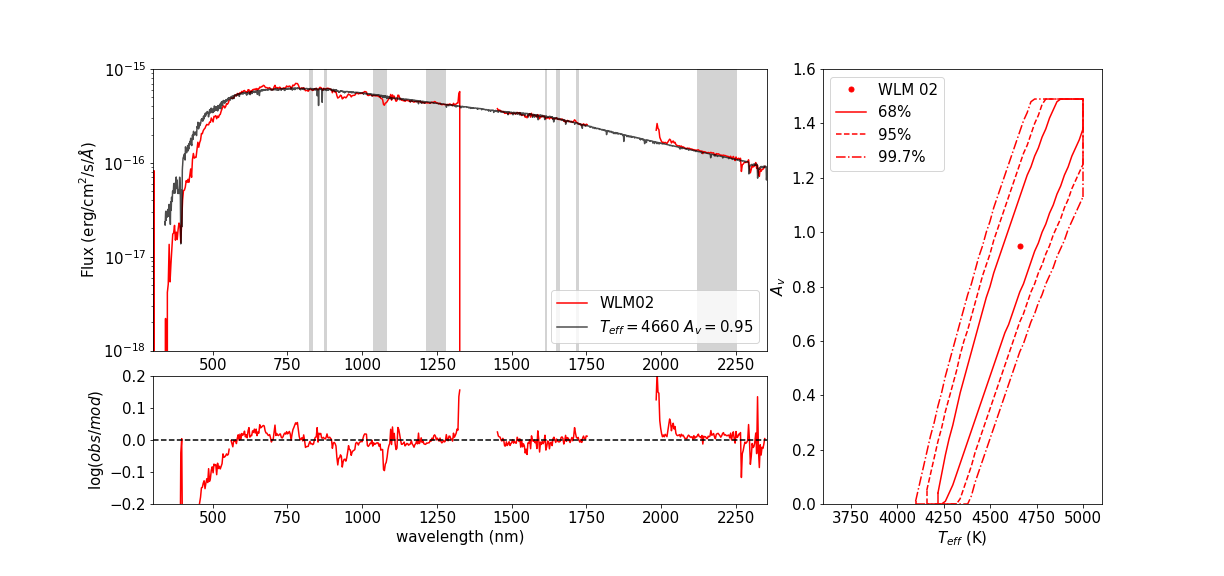}
\caption{Same as Fig.~\ref{fig:wlm1} but for WLM 02.}
\label{fig:wlm2}
\end{figure*}

\begin{figure*}
\centering
\includegraphics[width=1.\linewidth]{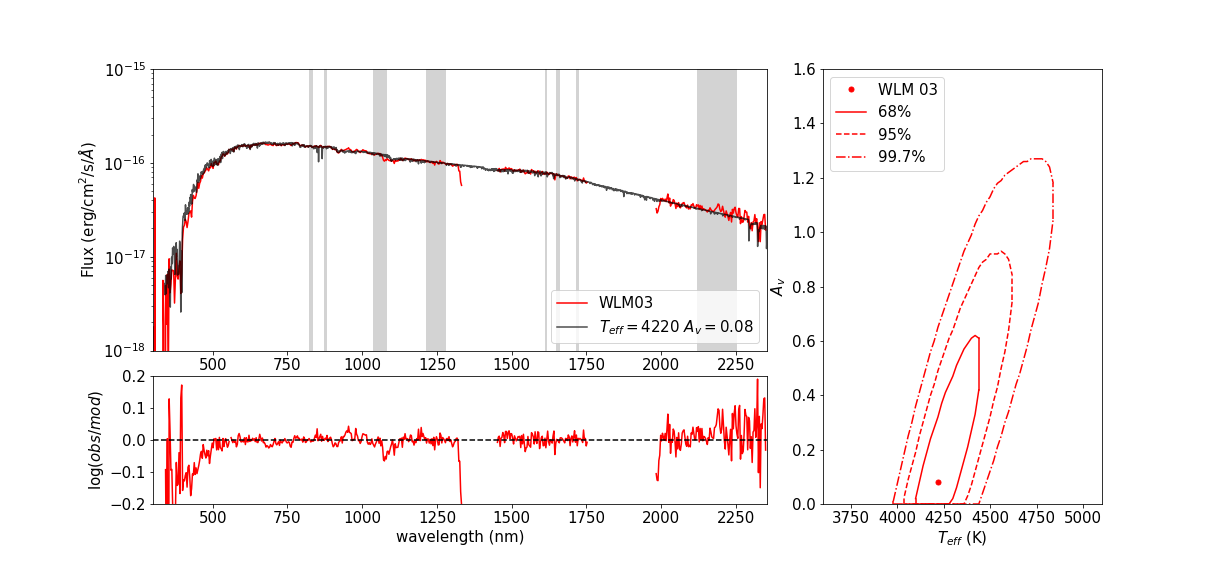}
\caption{Same as Fig.~\ref{fig:wlm1} but for WLM 03.}
\label{fig:wlm3}
\end{figure*}

\begin{figure*}
\centering
\includegraphics[width=1.\linewidth]{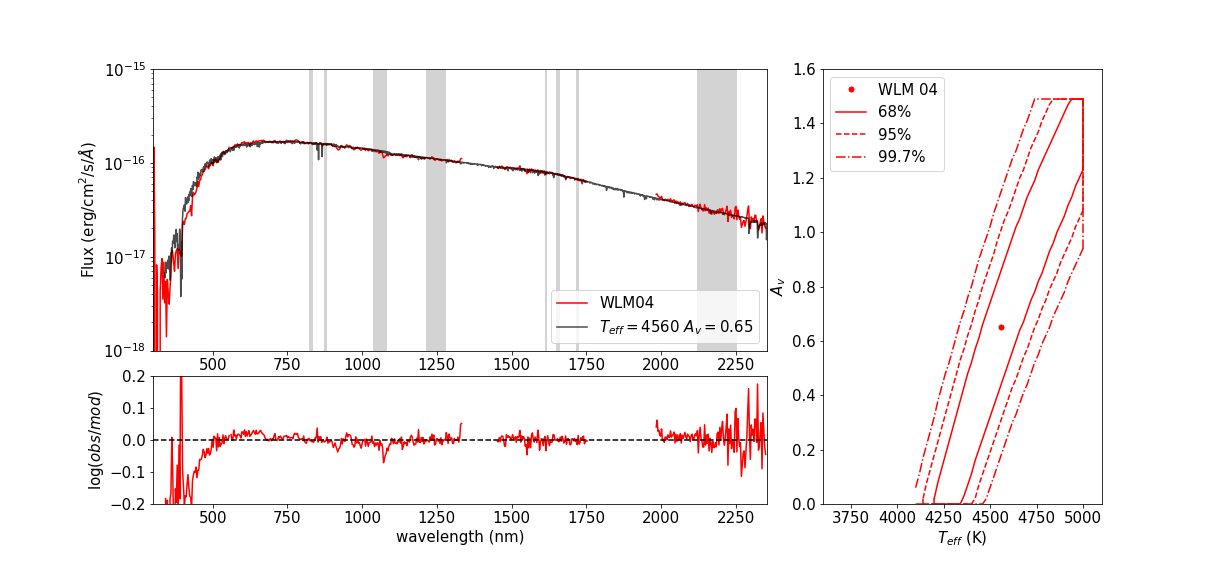}
\caption{Same as Fig.~\ref{fig:wlm1} but for WLM 04.}
\label{fig:wlm4}
\end{figure*}

\begin{figure*}
\centering
\includegraphics[width=1.\linewidth]{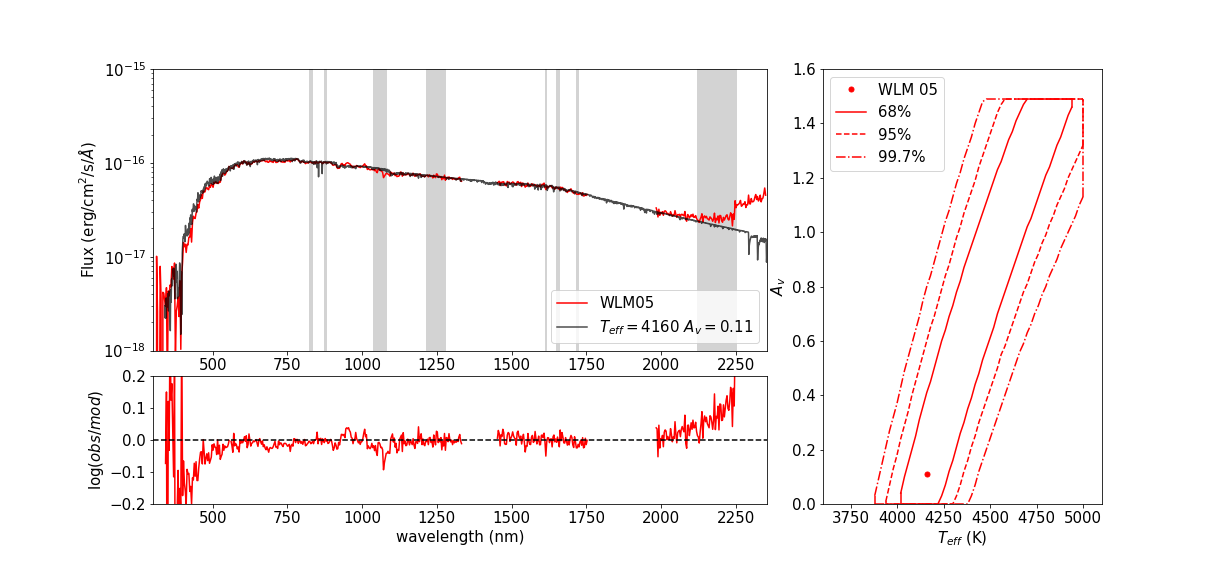}
\caption{Same as Fig.~\ref{fig:wlm1} but for WLM 05.}
\label{fig:wlm5}
\end{figure*}

\begin{figure*}
\centering
\includegraphics[width=1.\linewidth]{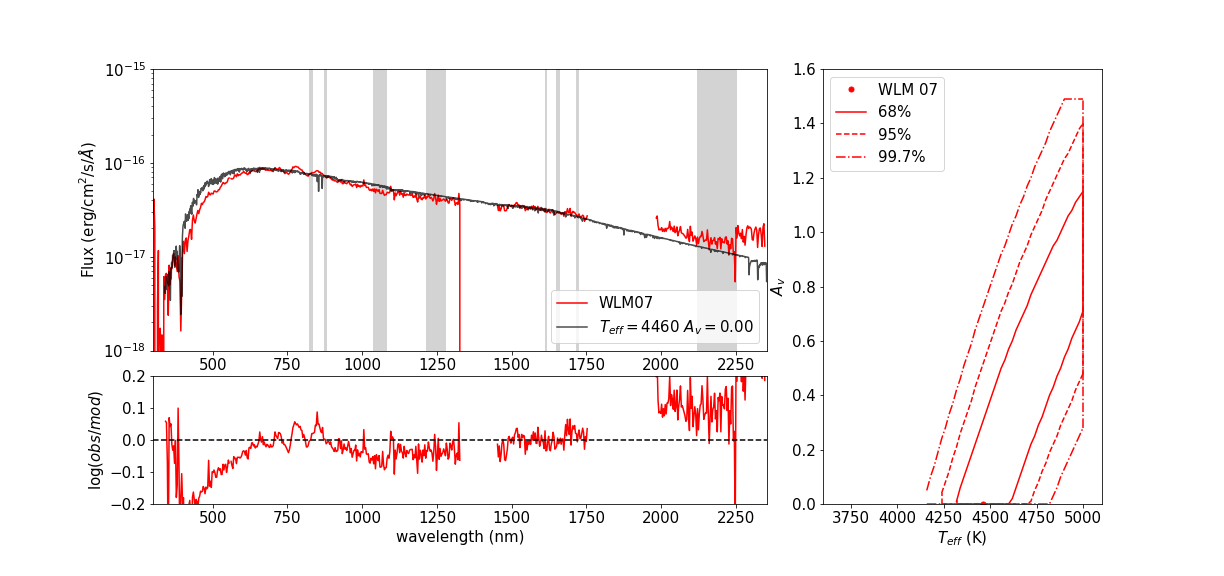}
\caption{Same as Fig.~\ref{fig:wlm1} but for WLM 07. This star is the problematic case and we suspect it is not part of the galaxy WLM.}
\label{fig:wlm7}
\end{figure*}

\begin{figure*}
\centering
\includegraphics[width=1.\linewidth]{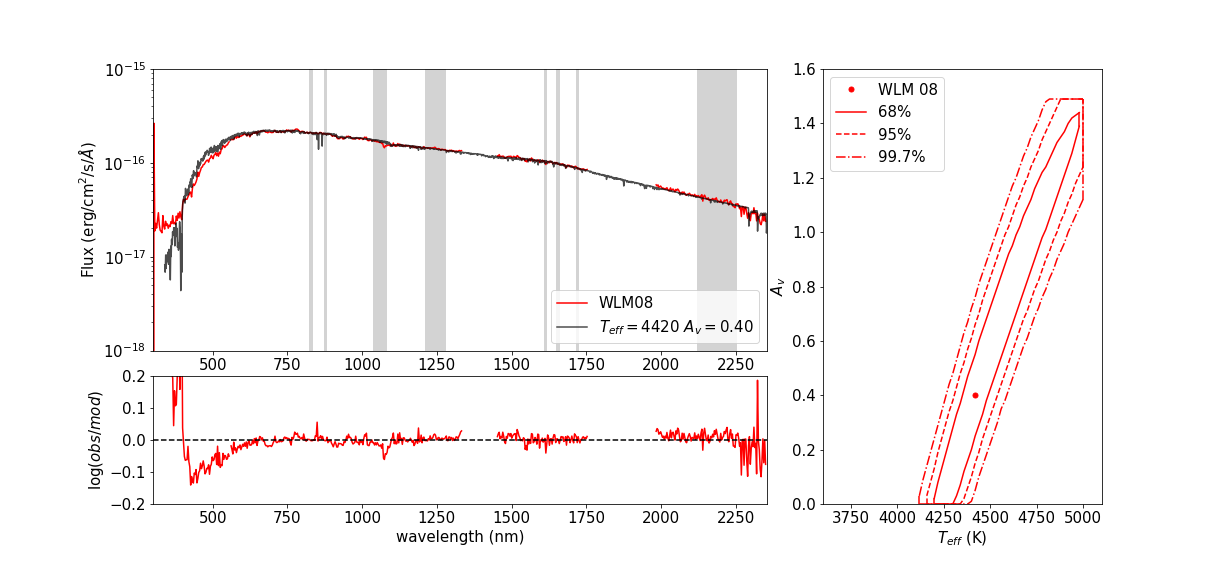}
\caption{Same as Fig.~\ref{fig:wlm1} but for WLM 08.}
\label{fig:wlm8}
\end{figure*}

\begin{figure*}
\centering
\includegraphics[width=1.\linewidth]{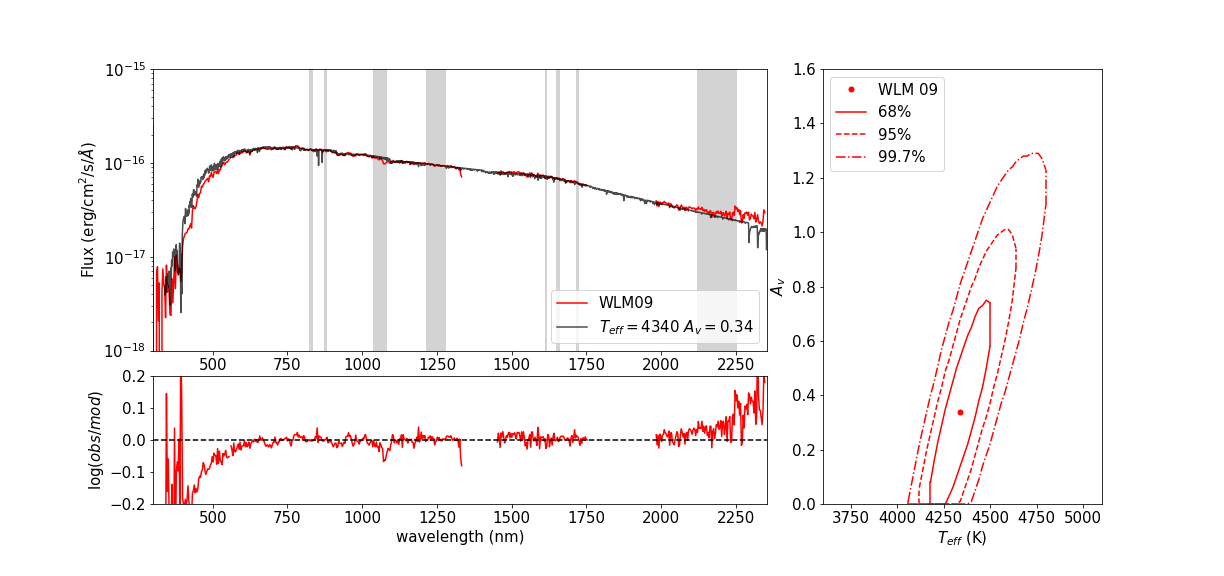}
\caption{Same as Fig.~\ref{fig:wlm1} but for WLM 09.}
\label{fig:wlm9}
\end{figure*}

\begin{figure*}
\centering
\includegraphics[width=1.\linewidth]{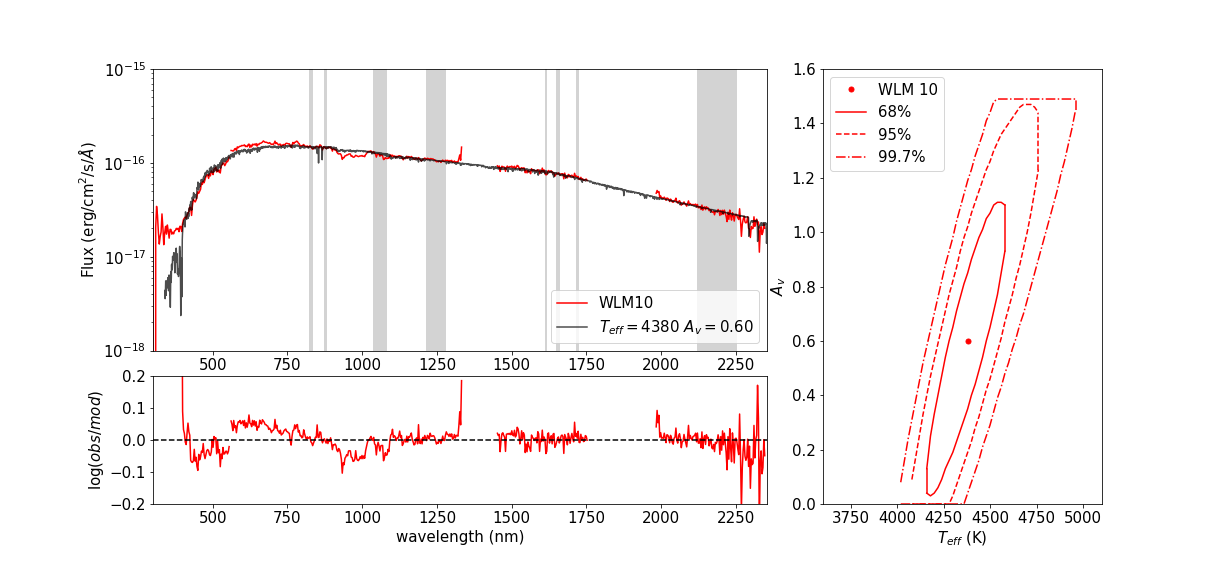}
\caption{Same as Fig.~\ref{fig:wlm1} but for WLM 10. }
\label{fig:wlm10}
\end{figure*}

\begin{figure*}
\centering
\includegraphics[width=1.\linewidth]{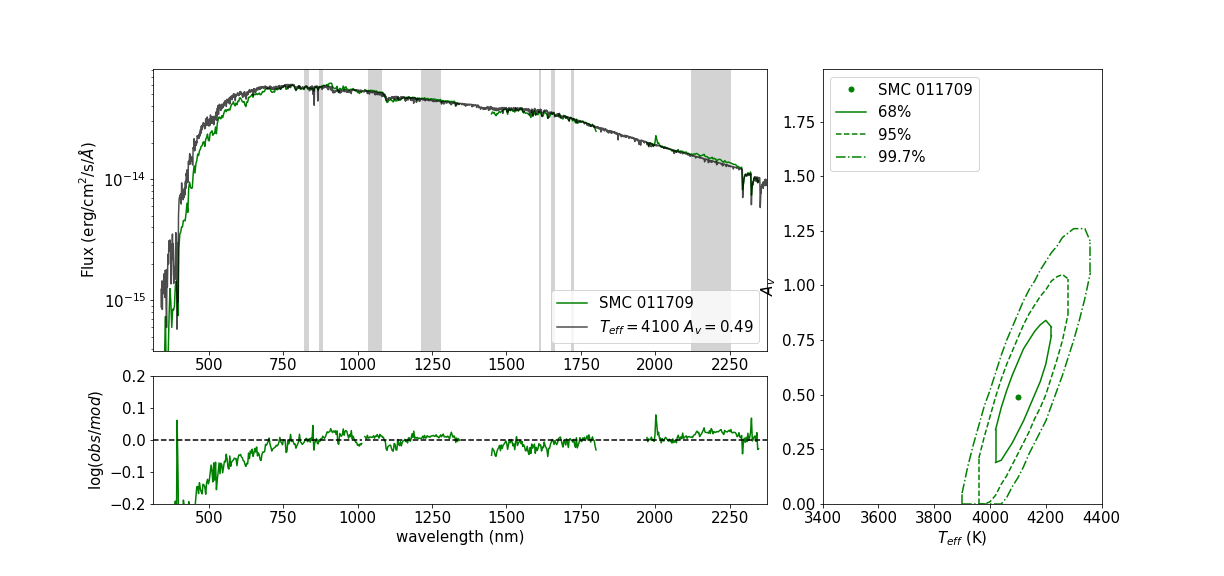}
\caption{Three panels showing the results of the analysis for the SMC 011709. \textit{Upper left:} Shows the smoothed data (green) and best fitted MARCS model (black), while the SED regions for the analysis are shown in gray. \textit{Lower left:} The residuals of the fit. \textit{Right:} shows the 68\%, 95\% and 99.7\% confidence contours for the best fitted effective temperature and extinction. }
\label{fig:smc1}
\end{figure*}

\begin{figure*}
\centering
\includegraphics[width=1.\linewidth]{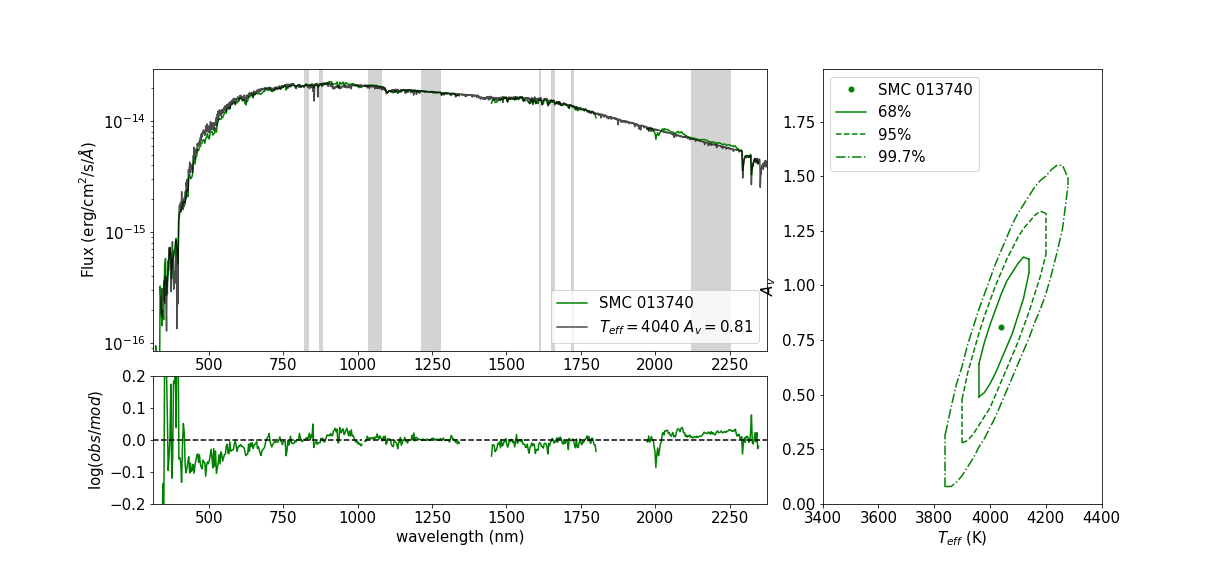}
\caption{Same as Fig.~\ref{fig:smc1} but for SMC 013740.}
\label{fig:smc2}
\end{figure*}

\begin{figure*}
\centering
\includegraphics[width=1.\linewidth]{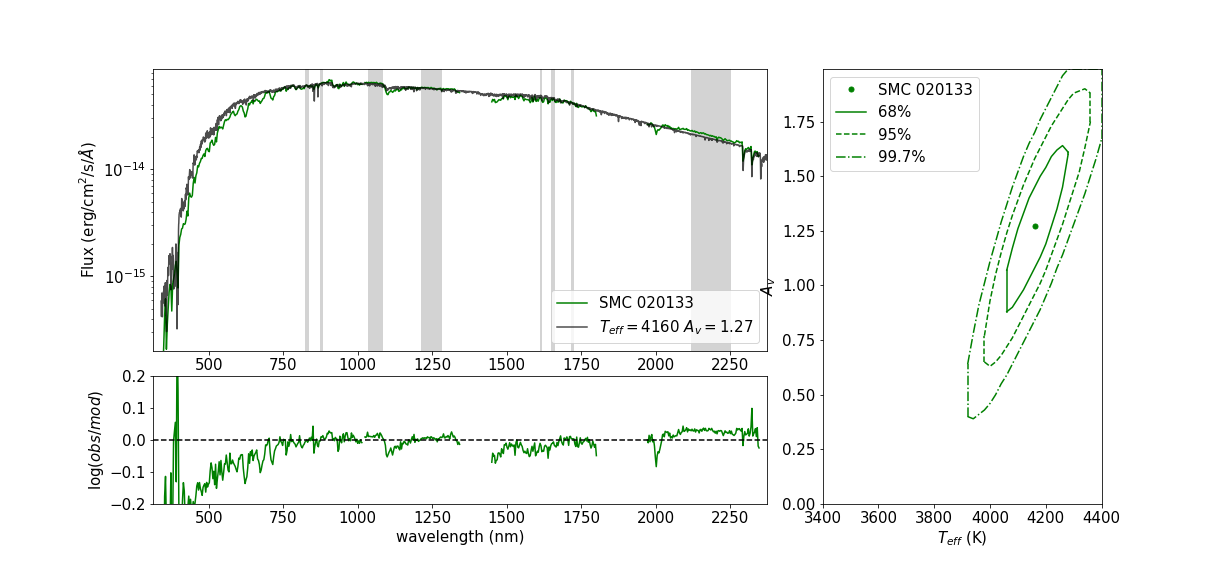}
\caption{Same as Fig.~\ref{fig:smc1} but for SMC 020133. }
\label{fig:smc3}
\end{figure*}

\begin{figure*}
\centering
\includegraphics[width=1.\linewidth]{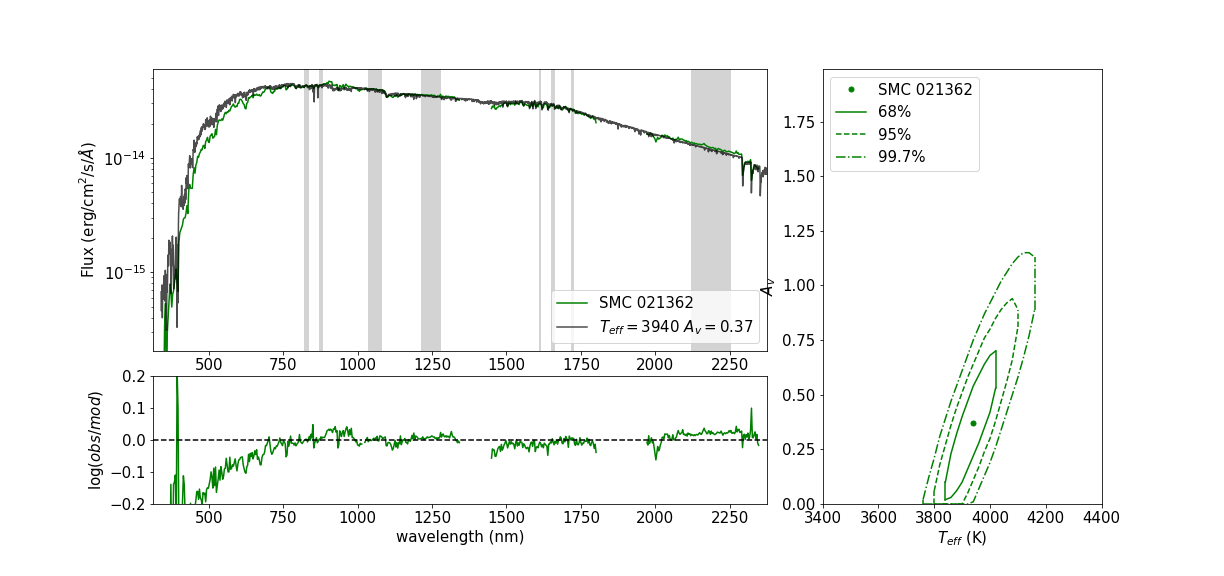}
\caption{Same as Fig.~\ref{fig:smc1} but for SMC 021362.}
\label{fig:smc4}
\end{figure*}

\begin{figure*}
\centering
\includegraphics[width=1.\linewidth]{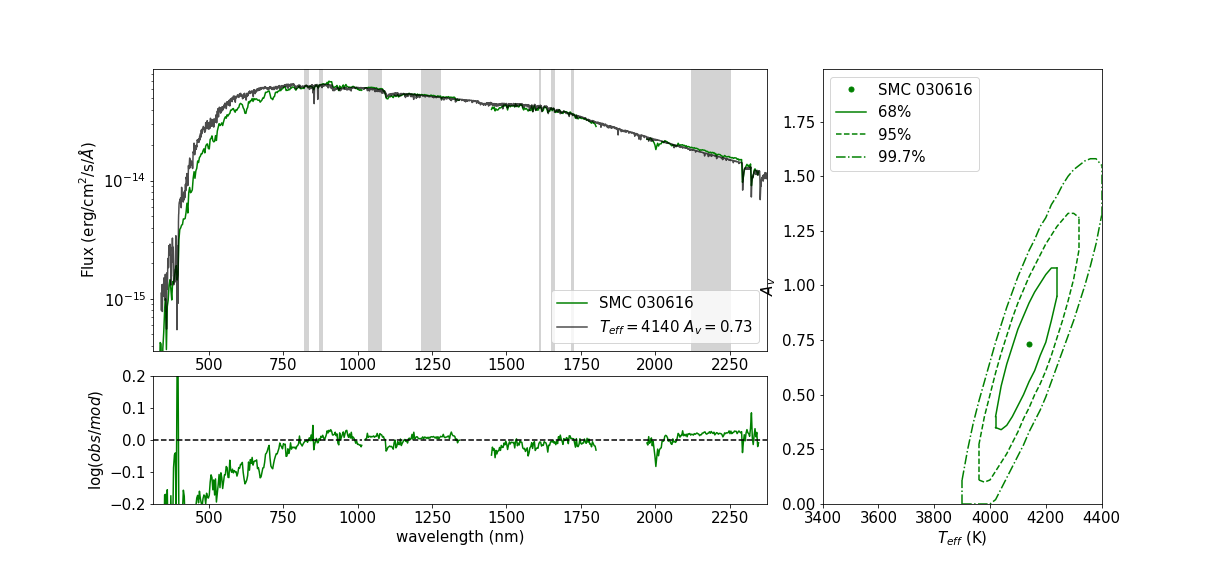}
\caption{Same as Fig.~\ref{fig:smc1} but for SMC 030616. }
\label{fig:smc5}
\end{figure*}

\begin{figure*}
\centering
\includegraphics[width=1.\linewidth]{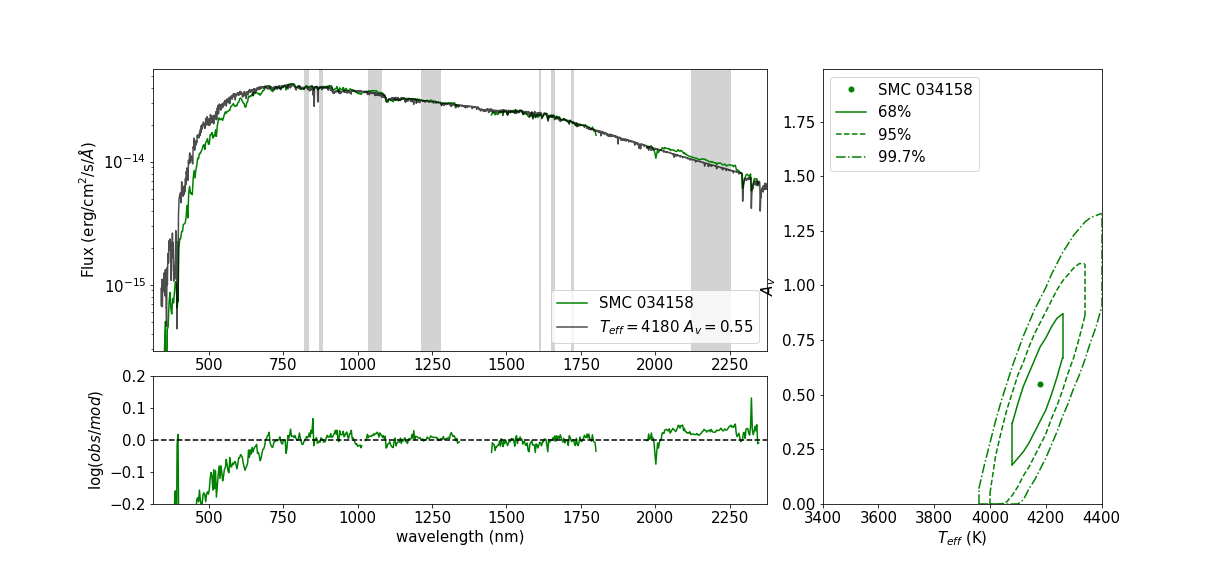}
\caption{Same as Fig.~\ref{fig:smc1} but for SMC 034158. }
\label{fig:smc6}
\end{figure*}

\begin{figure*}
\centering
\includegraphics[width=1.\linewidth]{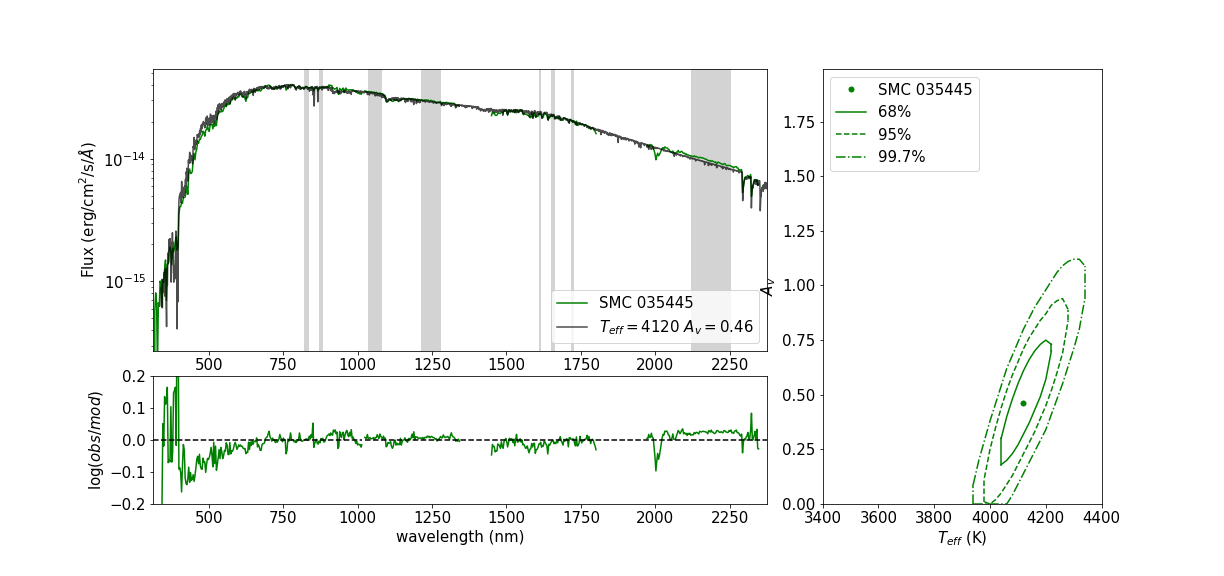}
\caption{Same as Fig.~\ref{fig:smc1} but for SMC 035445. }
\label{fig:smc7}
\end{figure*}

\begin{figure*}
\centering
\includegraphics[width=1.\linewidth]{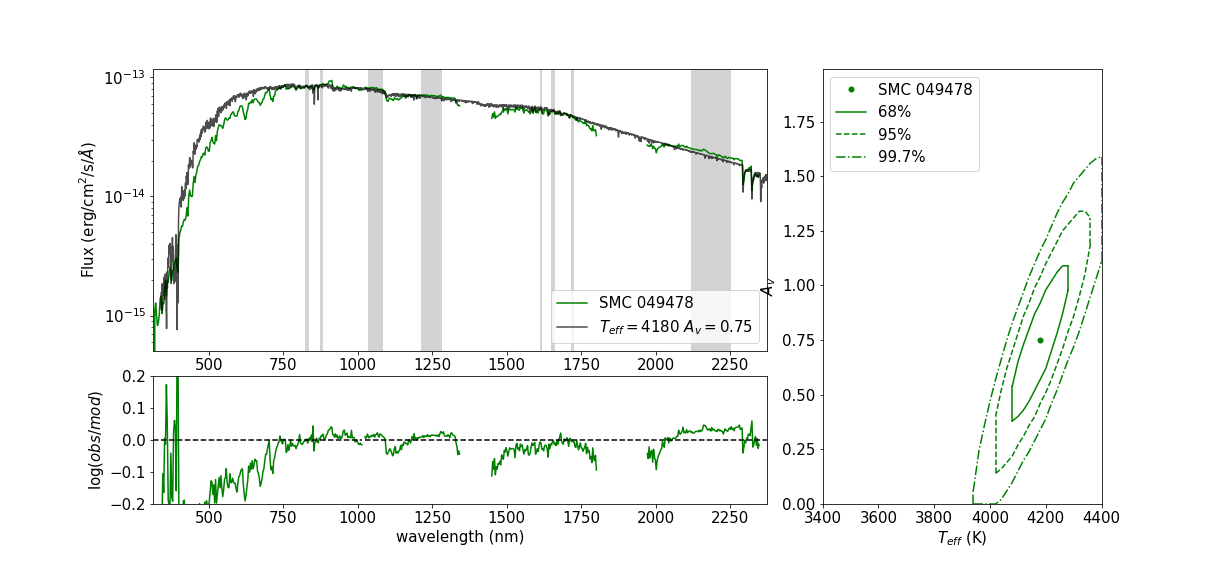}
\caption{Same as Fig.~\ref{fig:smc1} but for SMC 049478. }
\label{fig:smc8}
\end{figure*}

\begin{figure*}
\centering
\includegraphics[width=1.\linewidth]{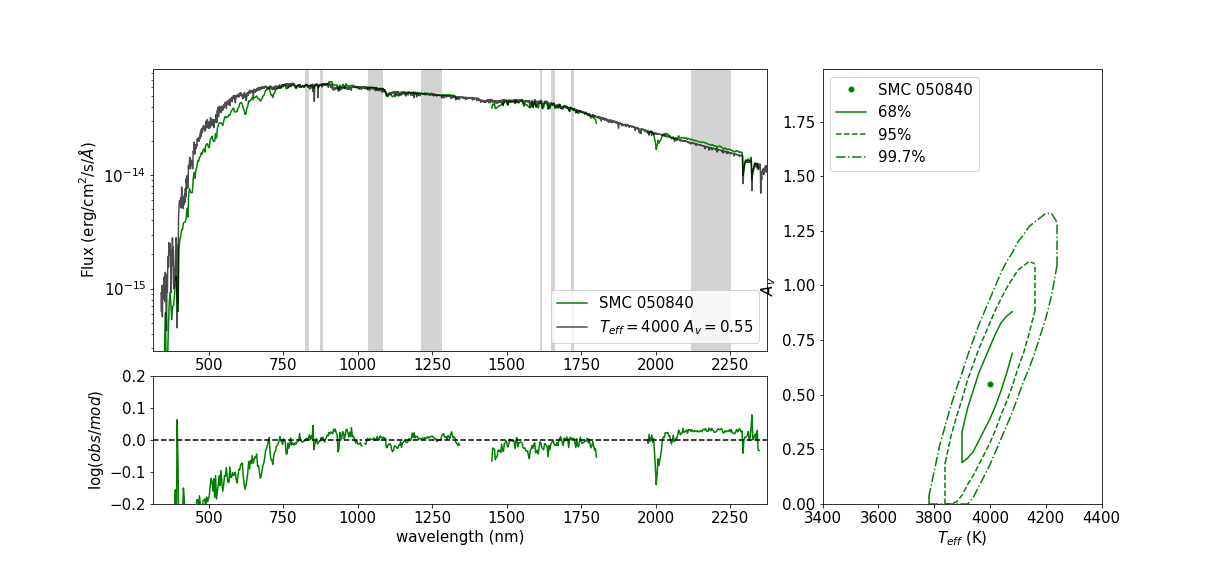}
\caption{Same as Fig.~\ref{fig:smc1} but for SMC 050840. }
\label{fig:smc9}
\end{figure*}

\begin{figure*}
\centering
\includegraphics[width=1.\linewidth]{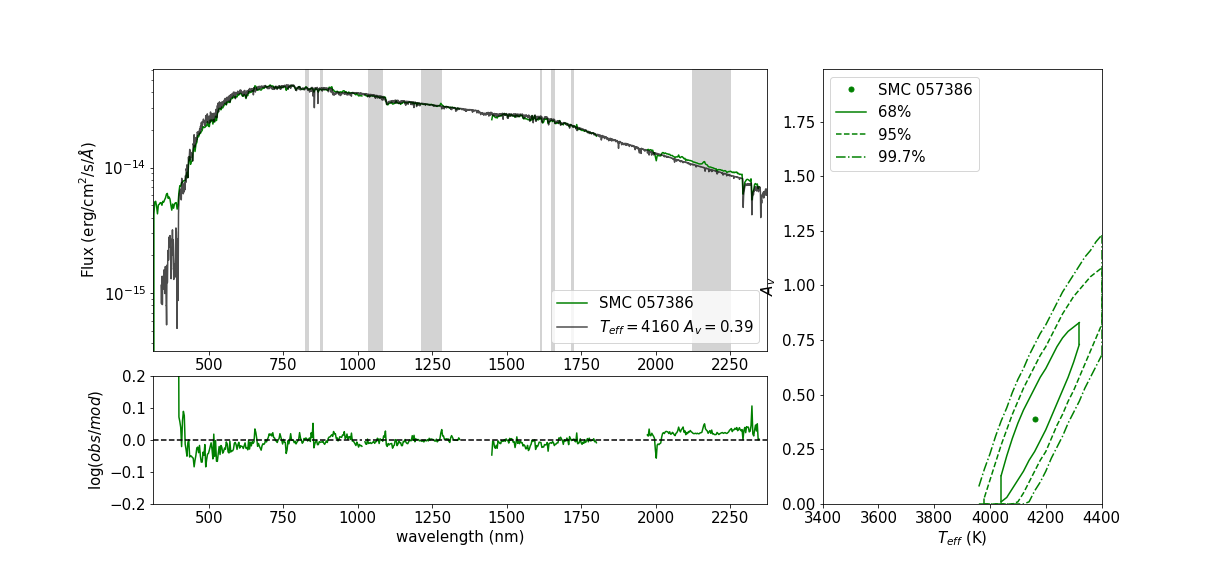}
\caption{Same as Fig.~\ref{fig:smc1} but for SMC 057386. }
\label{fig:smc10}
\end{figure*}

\begin{figure*}
\centering
\includegraphics[width=1.\linewidth]{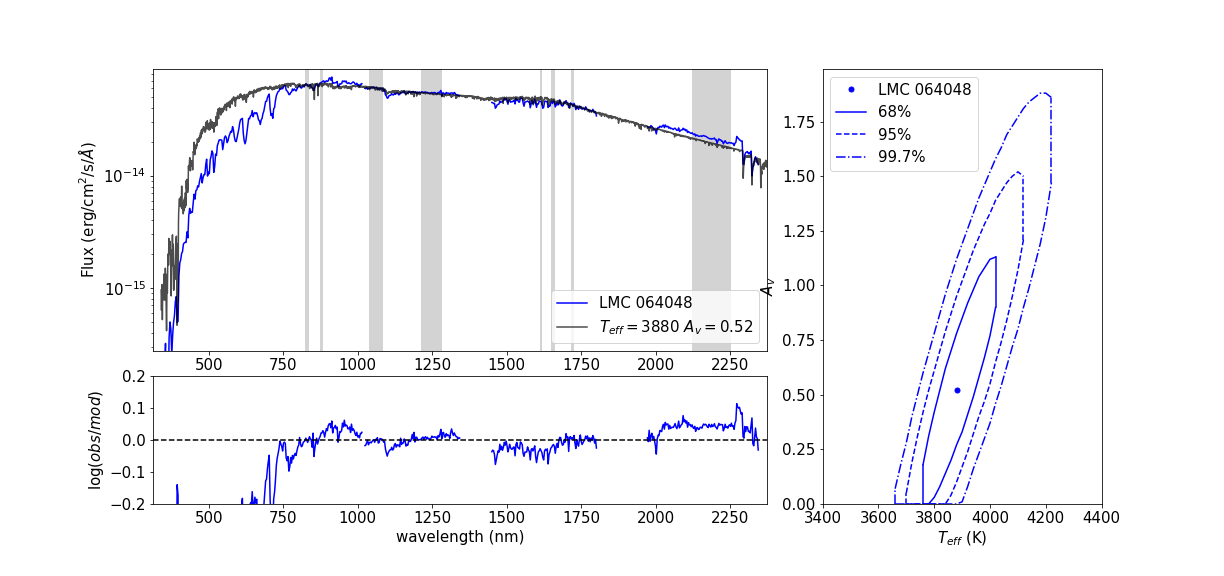}
\caption{Three panels showing the results of the analysis at the LMC 064048. \textit{Upper left:} Shows the smoothed data (blue) and best fitted MARCS model (black), while the SED regions for the analysis are shown in gray. \textit{Lower left:} The residuals of the fit. \textit{Right:} shows the 68\%, 95\% and 99.7\% confidence contours for the best fitted effective temperature and extinction. }
\label{fig:lmc1}
\end{figure*}

\begin{figure*}
\centering
\includegraphics[width=1\linewidth]{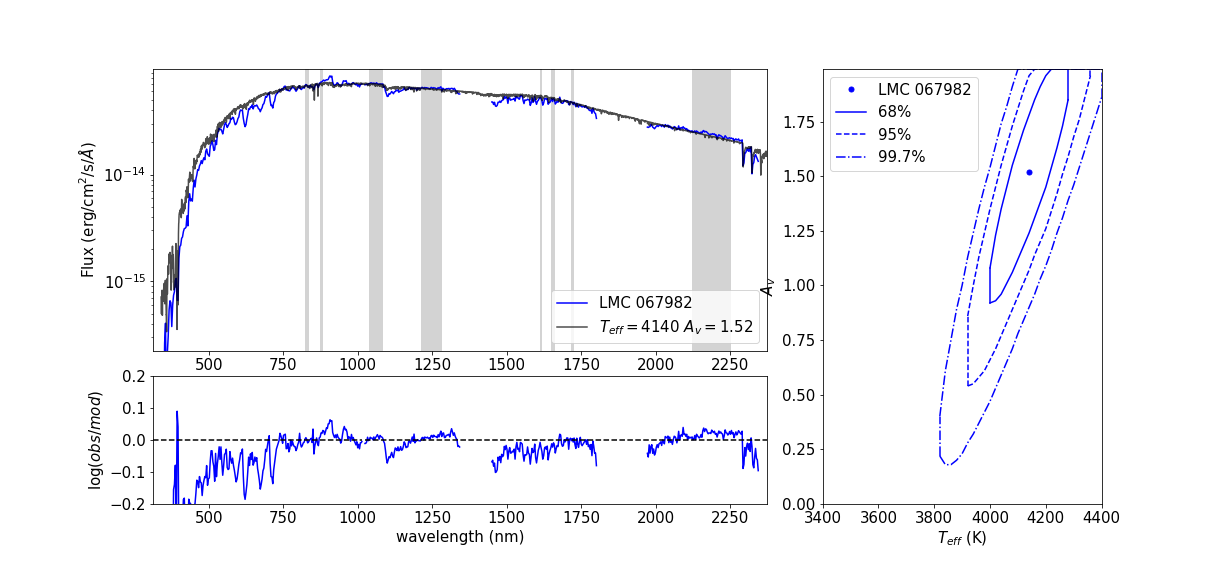}
\caption{Same as Fig.~\ref{fig:lmc1} but for LMC 067982. }
\label{fig:lmc2}
\end{figure*}

\begin{figure*}
\centering
\includegraphics[width=1\linewidth]{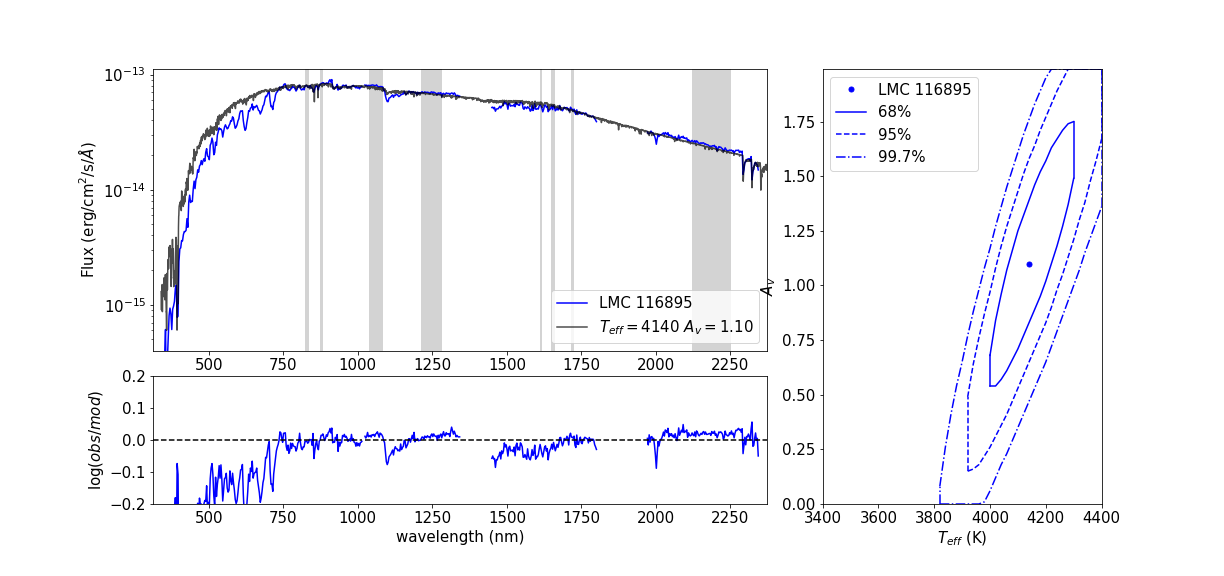}
\caption{Same as Fig.~\ref{fig:lmc1} but for LMC 116895. }
\label{fig:lmc3}
\end{figure*}

\begin{figure*}
\centering
\includegraphics[width=1\linewidth]{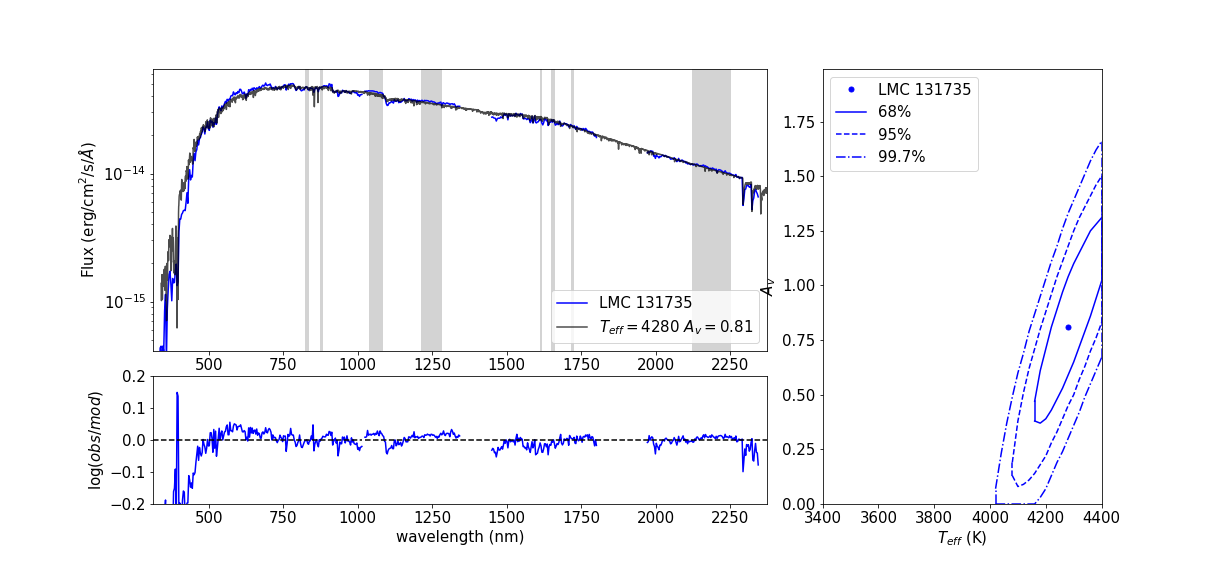}
\caption{Same as Fig.~\ref{fig:lmc1} but for LMC 131735. }
\label{fig:lmc4}
\end{figure*}

\begin{figure*}
\centering
\includegraphics[width=1\linewidth]{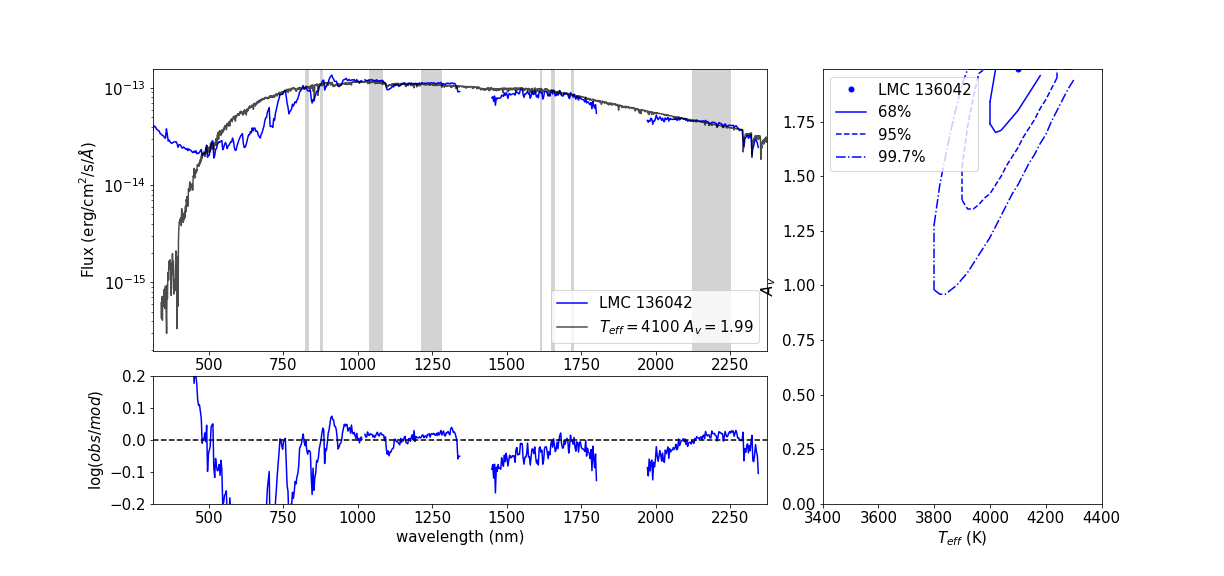}
\caption{Same as Fig.~\ref{fig:lmc1} but for LMC 136042. This is a problematic case that can have a near blue star  that contaminates the spectrum. }
\label{fig:lmc5}
\end{figure*}

\begin{figure*}
\centering
\includegraphics[width=1\linewidth]{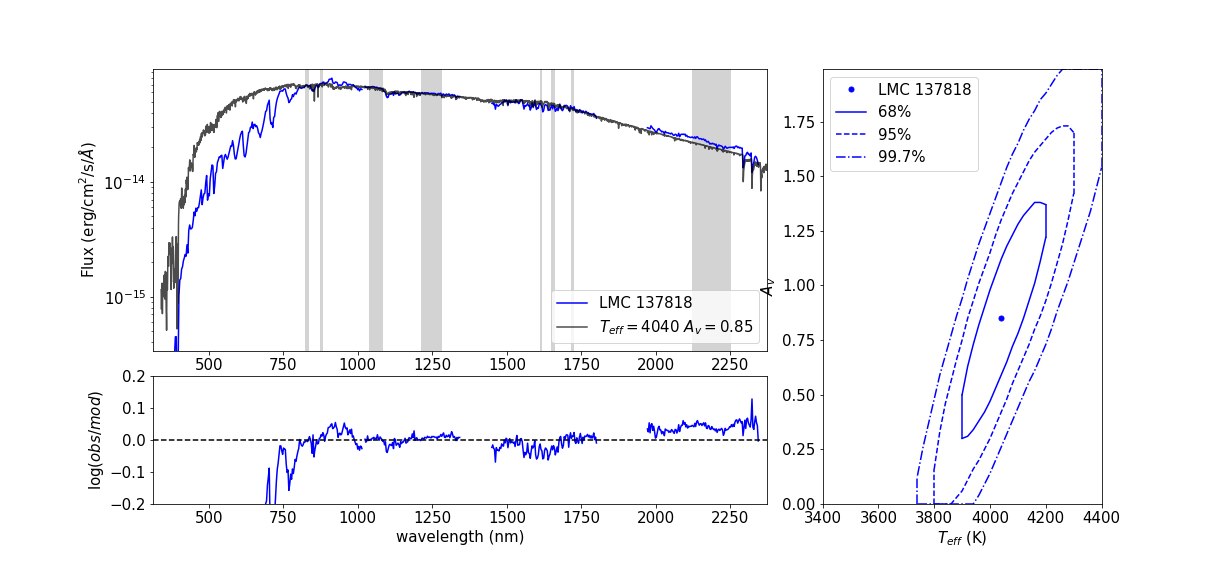}
\caption{Same as Fig.~\ref{fig:lmc1} but for LMC 137818.}
\label{fig:lmc6}
\end{figure*}

\begin{figure*}
\centering
\includegraphics[width=1\linewidth]{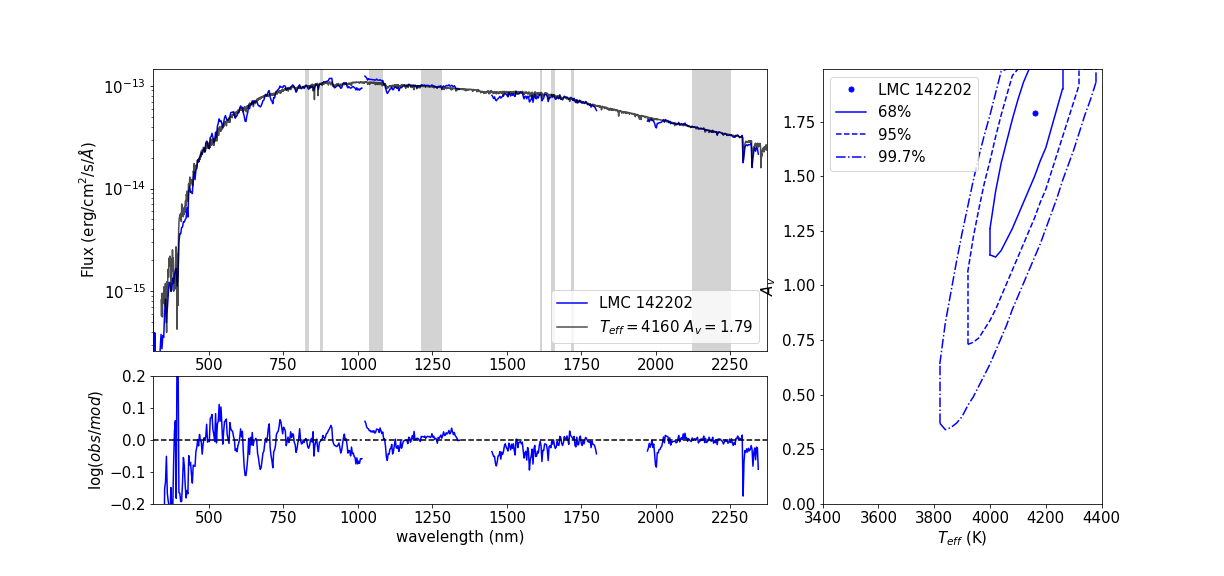}
\caption{Same as Fig.~\ref{fig:lmc1} but for LMC 142202. }
\label{fig:lmc7}
\end{figure*}

\begin{figure*}
\centering
\includegraphics[width=1\linewidth]{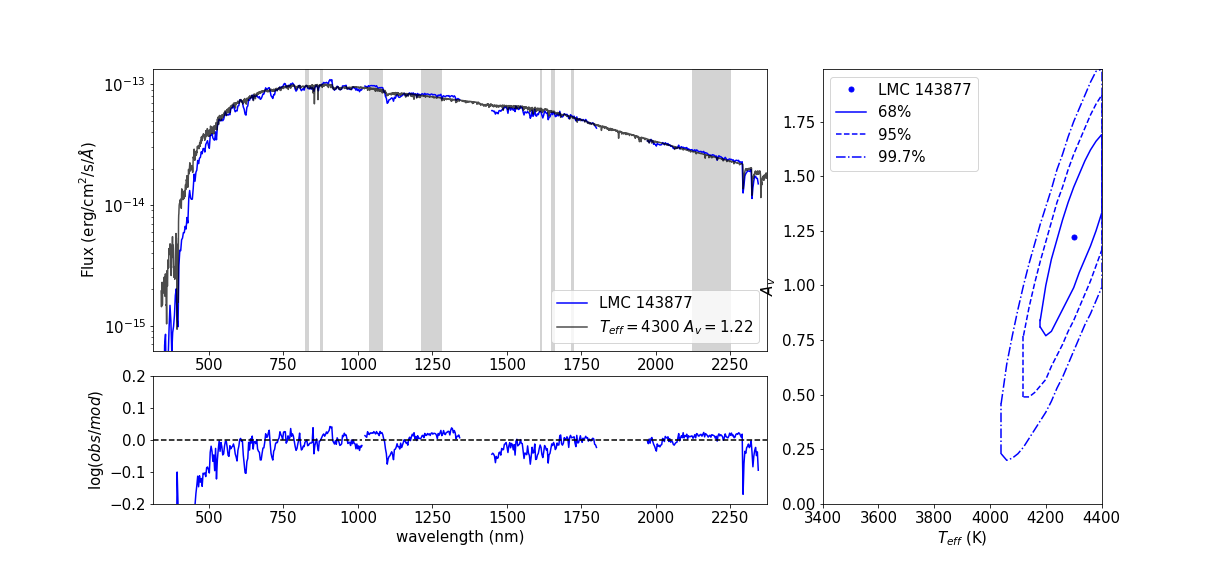}
\caption{Same as Fig.~\ref{fig:lmc1} but for LMC 143877. }
\label{fig:lmc8}
\end{figure*}

\begin{figure*}
\centering
\includegraphics[width=1\linewidth]{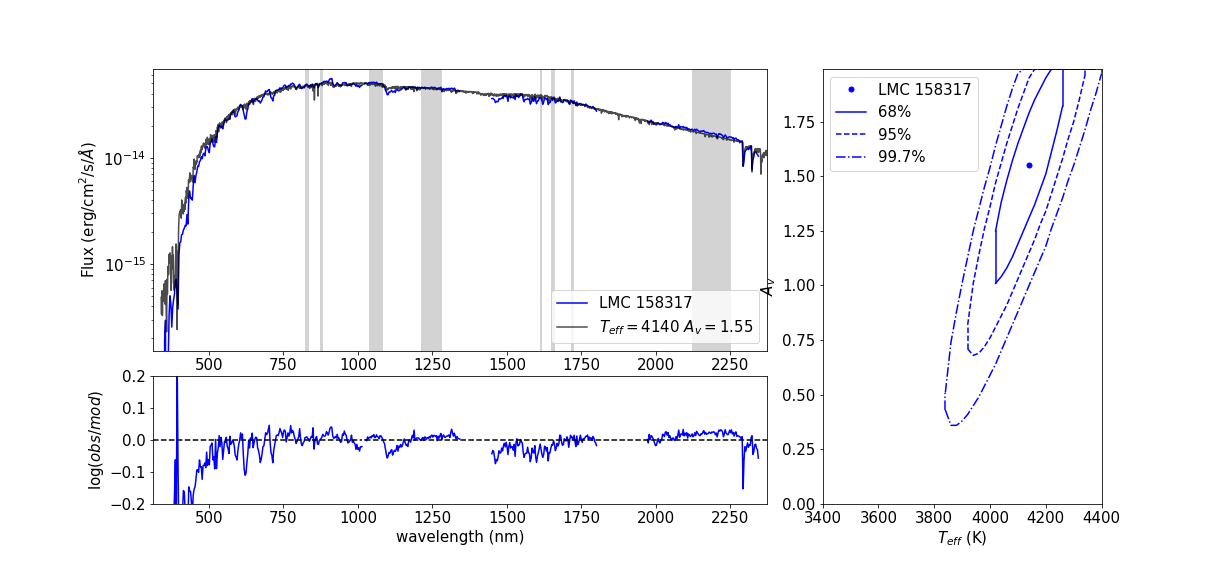}
\caption{Same as Fig.~\ref{fig:lmc1} but for LMC 158317.}
\label{fig:lmc9}
\end{figure*}
\end{appendices}
\bsp	
\label{lastpage}
\end{document}